\documentclass[sigconf,natbib=true,nonacm]{acmart}
\usepackage{graphicx}
\usepackage[english]{babel}

\author{Haya Nachimovsky}
\email{haya.nac@campus.technion.ac.il}
\affiliation{%
  \institution{Technion}
  \city{Haifa}
  \country{Israel}
  }

\author{Moshe Tennenholtz}
\email{moshet@technion.ac.il}
\affiliation{%
  \institution{Technion}
  \city{Haifa}
  \country{Israel}
  }

\author{Oren Kurland}
\email{kurland@technion.ac.il}
\affiliation{%
  \institution{Technion}
  \city{Haifa}
  \country{Israel}
  }
\title{A Multi-Agent Perspective on Modern Information Retrieval}

\usepackage{subcaption}
\usepackage{bbm}
\usepackage{algorithm}
\usepackage{algorithmic}
\usepackage{amsmath}
\usepackage{pifont}
\usepackage{graphicx}
\usepackage{xspace}
\usepackage{epsfig}
\usepackage{url}
\usepackage{booktabs}
\usepackage{balance}
\usepackage{thmtools}
\usepackage{amsfonts} 
\usepackage{amsthm}
\usepackage{multirow}
\usepackage{tikz}

\newcommand{\myparagraph}[1]{\vspace{0.6\baselineskip}\noindent{\textbf{#1}}.~}

\newcommand{\omt}[1]{}
\newcommand{\firstmention}[1]{{\bf #1}}

\newcommand{\ltr}{LTR\xspace}

\newcommand{\llmNoX}{LLM}

\newcommand{\llm}{\llmNoX\xspace}

\newcommand{\lambdamart}{LambdaMART\xspace}
\newcommand{\bert}{BERT\xspace}

\newcommand{\bm}{BM25\xspace}

\newcommand{\tfidf}{TF.IDF\xspace}

\newcommand{\ltrWithX}{LTR\xspace}

\newcommand{\neu}{NEU\xspace}

\newcommand{\single}{S\xspace}

\newcommand{\multi}{M\xspace}

\newcommand{\NivDataset}{\single-\neu}

\newcommand{\GregDataset}{\single-\ltrWithX}

\newcommand{\MultiB}{\multi-\ltrWithX}

\newcommand{\MultiD}{\multi-\neu}

\newcommand{\lexicalagent}{Lexical\xspace}
\newcommand{\semanticagent}{Semantic\xspace}
\newcommand{\llmagent}{LLM\xspace}

\newcommand{\efive}{E5\xspace}
\newcommand{\contriever}{Contriever\xspace}

\newcommand{\llama}{Llama\xspace}
\newcommand{\llamaWithVersion}{Llama 3.1\xspace}
\newcommand{\gemma}{Gemma\xspace}
\newcommand{\gemmaWithVersion}{Gemma 2\xspace}
\newcommand{\gpt}{GPT\xspace}
\newcommand{\gptWithVersion}{GPT-3.5\xspace}

\newcommand{\human}{human\xspace}

\newcommand{\mixed}{mixed\xspace}

\newcommand{\interpolation}{$\lambda$\xspace}

\newcommand{\interpolationMath}{\lambda}

\newcommand{\all}{All}
\newcommand{\better}{Better}
\newcommand{\best}{Best}

\newcommand{\nlithreshold}{$\eta$\xspace}

\newcommand{\nlithresholdMath}{\eta}
\newcommand{\numhighestranked}{$m$\xspace}

\newcommand{\numhighestrankedMath}{m}

\newcommand{\feedbackAll}{$F_{all}$}
\newcommand{\feedbackPair}{$F_{pair}$}

\newcommand{\nocopyMath}{NoCopy}
\newcommand{\nocopy}{$\nocopyMath$}

\newcommand{\withoutnocopy}{$\neg\nocopyMath$}

\newcommand{\botagent}{document agent\xspace}

\newcommand{\botagents}{document agents\xspace}

\newcommand{\purehumancorpus}{human\xspace}
\newcommand{\purellmcorpus}{LLM\xspace}
\newcommand{\mixedcorpus}{mixed\xspace}
\newcommand{\agentcorpus}{corpus\xspace}

\newcommand{\source}{s}
\newcommand{\target}{t}
\newcommand{\SrcSentenceMath}{g_{\text{\source}}}
\newcommand{\TargetSentenceMath}{g_{\text{\target}}}
\newcommand{\SrcSentence}{${\SrcSentenceMath}$\xspace}
\newcommand{\TargetSentence}{${\TargetSentenceMath}$\xspace}

\newcommand{\QryTerm}{QT}
\newcommand{\QryTermSrc}{$\QryTerm_{\source}$}
\newcommand{\QryTermTarget}{$\QryTerm_{\target}$}
\newcommand{\QryTermSrcMath}{\QryTerm_{\source}}
\newcommand{\QryTermTargetMath}{\QryTerm_{\target}}

\newcommand{\SimTop}{ST}
\newcommand{\SimSrcTop}{$\SimTop_{\source}$}
\newcommand{\SimTargetTop}{$\SimTop_{\target}$}
\newcommand{\SimSrcTopMath}{\SimTop_{\source}}
\newcommand{\SimTargetTopMath}{\SimTop_{\target}}

\newcommand{\highlight}[1]{{\em#1}}





\begin{abstract}
    The rise of large language models (LLMs) has introduced a new era in information retrieval (IR), where queries and documents that were once assumed to be generated exclusively by humans can now also be created by automated agents. These agents can formulate queries, generate documents, and perform ranking. This shift challenges some long-standing IR paradigms and calls for a reassessment of both theoretical frameworks and practical methodologies.
We advocate for a multi-agent perspective to better capture the complex interactions between query agents, document agents, and ranker agents. 
Through empirical exploration of various multi-agent retrieval settings, we reveal the significant impact of these interactions on system performance.
Our findings underscore the need to revisit classical IR paradigms and develop new frameworks for more effective modeling and evaluation of modern retrieval systems.
\end{abstract}

\begin{document}
\maketitle



\section{Introduction}
There is a growing body of work and discussion on the potential impact
of large language models (LLMs) on the information retrieval (IR)
field \cite{Chen+al:21a,White:25a,Zhai:24a,Zhu+al:24a}. 
A prominent theme is using LLM-based agents to assist the user in information-seeking tasks~\cite{Chen+al:21a,Shah+White:24a,White:25a,Zhai:24a,Zhang+al:24a}. 
For example, these agents, henceforth referred to as {\em query agents}, can help formulate queries representing users’ information needs~\cite{Chen+al:21a,Shah+White:24a,White:25a,Zhai:24a}.

There are two additional types of agents which together with query
agents play a crucial role in modern retrieval
settings. The first are {\em document agents}: agents whose goal is to
assist in document generation and editing. Indeed, the proliferation
of LLM-based generated content is ever growing
\cite{ai-generated-social-media}. The second type of agents is that
used to induce ranking, specifically for ad hoc (query-based) retrieval which
is our focus in this paper. Herein, we refer to this type of agent as
{\em ranker agent}. Accordingly, in the ad hoc retrieval setting we
consider, documents in the corpus can be generated by humans or
document agents, queries representing users' information needs can be
generated by the users themselves or agents acting on their behalf\footnote{Agents can also have information needs, but this 
  is outside the scope of this paper.}, and the ranking mechanism can be an agent (e.g., LLM-based rankers
\cite{liang2022holistic,qin2024pairwise,zhuang2024setwise}).

We argue in this paper
that the mutual effects among the three types of agents just discussed --- query, document and ranker --- call for
re-consideration of
some classical retrieval paradigms and frameworks. Furthermore, there are far
reaching implications on evaluation of ad hoc retrieval. As we discuss below, we do not confine our call for arms for LLM-based agents only.

Consider, for example, the generative theory for relevance
\cite{Lavrenko+Croft:01a,lavrenko-rm-book}. The relevance assumption is
that there is a relevance language model\footnote{The specific language
  models were unigram.} that generates terms in the query and in
documents relevant to the information need it represents. Since
document and query agents need not be aligned --- e.g., they can
utilize completely different language models --- this generative
assumption does not necessarily hold. In fact, the potential
misalignment between document and query agents is conceptually
reminiscent of the state-of-affairs in cross-lingual retrieval
\cite{Lavrenko+al:02a}. The misalignment between the ranker agent and document agent can also have significant effect. A case in point, recent work \cite{neural-llm-bias,dai2024cocktail}
shows that LLM-based rankers are biased in favor of LLM-based
generated content.

Another important aspect of multi-agent retrieval settings is ranking incentives.
In competitive search
settings \cite{kurland_competitive_2022}, some document authors are
incentivized to have their documents highly ranked for specific
queries (e.g., queries of commercial intent). They often modify documents in response to induced ranking so as to improve their future ranking.
There is recent work on devising document agents that automatically modify documents
for rank promotion, while preserving document quality
\cite{Greg-bot,Niv}. Commonly used modification strategies were shown
to lead to herding effects not only of human authors
\cite{Nimrod,Greg-Herding}, but also of LLM-based document agents
\cite{Lemss}. The resultant corpus effects (e.g., reduced topical
diversity) are unwarranted~\cite{Nimrod}. Devising ranking functions (i.e., ranker
agents) that account for both search effectiveness and long-term
corpus effects is an open challenge
\cite{kurland_competitive_2022}. Devising query agents that operate in 
the competitive search setting is an additional important research
question we discuss.

The foundational Cranfield evaluation paradigm \cite{cranfield}, which
is the underpinning of many evaluation practices in IR (e.g., those
used in TREC \cite{Harman+Voorhees:06a}), is based on using a test
collection that includes documents, information needs (and queries
representing them) and relevance judgments. Evaluating search
effectiveness in the multi-agent search setting we discuss poses
several evaluation challenges. First, creating a test collection
composed of documents generated by a representative variety of
document agents is a challenge given the rapid emergence of new LLM
types and technologies. Such collections can become obsolete very
quickly. Second, a similar challenge emerges due to query
agents which generate queries. Different agent technologies can result
in completely different types of queries. Additional complexity is introduced by the need to account for
competitive search conditions \cite{kurland_competitive_2022} as those
mentioned above. Static collections cannot support evaluation of corpus effects driven by document agents responding to induced rankings \cite{kurland_competitive_2022}. Accounting for all these considerations together, we argue for the increasing importance of simulation-based evaluation in multi-agent retrieval settings (cf., \cite{Lemss}).

Our call for arms in this paper, about re-considering the fundamentals
of ad hoc retrieval in multi-agent settings, is composed of two
parts. 
In the first part, we present our perspectives about the setting and potential future research directions it gives rise to.
In the
second part, we present an in-depth empirical illustration of the
multi-agent retrieval setting and its potential consequences. We use
three different types of approaches to devise document, ranker and
query agents: lexical (\tfidf-based), semantic (embedding-based), and
LLM-based. Some of these agent types are novel to this study.  Our
empirical findings clearly support our call for arms. For example,
when the query agent and ranker agent are of different types (e.g.,
one is semantic and the other is LLM-based), retrieval effectiveness
degrades with respect to having both of the same type. Additional example is the result of misalignment between the types of document and ranker agents. If both are LLM-based, but implemented with different LLMs, then the ability of the document agent to promote its document in rankings is reduced with respect to a setting where the same LLM is used for both the document and the ranker agents.

The paper is structured as follows. In Sections \ref{sec:rankAgent},
\ref{sec:query-agent} and \ref{sec:docAgent} we present our perspectives about
the ranker (agent), query (agent) and document (agent), respectively, in the multi-agent setting.
In
Section \ref{sec:evalConsider}, we discuss the evaluation implications of the multi-agent retrieval setting. Sections
\ref{sec:experimental-settings} 
and \ref{sec:evaluation} present an
empirical exploration of various multi-agent retrieval settings.

\section{The Ranker Agent}
\label{sec:rankAgent}
The ranker agent has to induce a ranking in response to a query as is
standard in ad hoc retrieval. We use the term ``ranker agent'' for
alignment with the terminology used for the document and query agents discussed below,
and to indicate that the ranking mechanism can be quite evolved: it can include a query understanding (a.k.a.,
query intent identification) module \cite{Jansen+al:07a}, utilize
pseudo relevance feedback \cite{Xu+Croft:96a}, or fuse lists retrieved
by different ranking functions \cite{Kurland+Culpepper:18a}.

Regardless of who generated the query --- i.e., the user
directly or a query agent --- the goal is to satisfy the
information need it represents. Herein we assume that the information
need is that of the user. Treatment of cases where agents have their
own information needs is outside the scope of this paper. Accordingly,
the probability ranking principle (PRP) \cite{Robertson:77a} remains
optimal in the multi-agent retrieval setting under the conditions
specified by Robertson \cite{Robertson:77a}: the relevance of
different documents is independent and users have the same utility
function. That is, the ranking should be based on the probability a
document is relevant to the query (or more precisely, the information
need it represents) where the probability is estimated using all
information available to the search system. The fundamental difference
between retrieval methods is the relevance estimate: it could be a
probability-based estimate (e.g., as in Okapi BM25
\cite{Robertson:93a}), or a proxy thereof (e.g., cosine between the
embedding vectors of the query and the document).

The different
frameworks and paradigms for relevance estimation can be significantly affected by
the fact that query and document agents might have generated the
query and the document, respectively. We next discuss a few examples of foundational relevance estimation paradigms that should be re-considered.


\myparagraph{The relevance model} The generative relevance assumption
is that terms in the query and in documents relevant to the query are
generated by the same language model \cite{Lavrenko+Croft:01a,lavrenko-rm-book}. The assumption, as well as the  relevance-model estimation approach, were based on unigram language models. Conceptually, the assumption should hold for more evolved language models.

However, since documents and queries can be generated by different types of agents, the original generative assumption, and its conceptual generalization just proposed, do not hold. This state-of-affairs is reminiscent of that in cross-lingual retrieval where queries and documents are written in different languages \cite{Lavrenko+al:02a,crossling}. Accordingly, re-visiting the generative theory to relevance is an interesting future direction to explore in multi-agent retrieval settings.

\myparagraph{The risk minimization framework} Lafferty and Zhai
\cite{Lafferty+Zhai:01a} describe the document authorship and
retrieval processes as follows. A document author selects a (language)
model from which she samples terms for the document she writes. The
user of the search system also selects a (language) model representing
the information need. Terms are sampled from the user model to
generate the query. To estimate document relevance, the query and
document models are compared. Since these are unigram language models \cite{Lafferty+Zhai:01a}, KL divergence and cross entropy are often used to compare the models.

Lafferty and Zhai's framework \cite{Lafferty+Zhai:01a} is conceptually
aligned at its most basic level with the multi-agent retrieval setting
we address here. The models selected by the document author and user
are now replaced with agents. The agents can use language models or
any other means to generate documents and queries. More importantly, due to the potential divergence
between the document and query agent, basic model comparison as that in
Lafferty and Zhai's framework \cite{Lafferty+Zhai:01a} does not
necessarily work. Extending the risk minimization framework to account for document and query agent type misalignment is an interesting future direction to explore.

\myparagraph{The axiomatic framework} The axiomatic framework for
retrieval \cite{Fang+Zhai:05a} has evolved to
also handle modern neural retrieval methods \cite{Cheng+Fang:20a}. The idea is to use a set of axioms to analyze existing
retrieval methods and to devise new ones.

In competitive search settings \cite{kurland_competitive_2022},
document authors might modify their documents so as to improve their
future ranking. This practice is often referred to as search engine
optimization (SEO) \cite{Gyongyi+Molina:05a}. There are recent
examples of document agents devised to modify document content for
improved ranking while maintaining document quality
\cite{Greg-bot,Niv}. As noted elsewhere
\cite{kurland_competitive_2022}, this reality undermines some of the
basic axioms for retrieval \cite{Fang+Zhai:05a}; e.g., that increased
query term occurence in a document should result in increased
retrieval score. Indeed, when document agents are ranking
incentivized, query term occurrence need not necessarily reflect pure
authorship considerations. An interesting future direction is to
revise the axiomatic approach to reflect that documents and queries can be generated by agents and these agents can be of completely different types (e.g., lexical vs. semantic), as we present in Section~\ref{sec:agents-implementation}.
  
\subsection{Devising Ranker Agents}
\label{sec:devising-ranker-agents}
The multi-agent retrieval settings brings about new challenges as
those mentioned above. At the same time, it gives rise to a plethora
of new opportunities, a few of which we now turn to discuss.

\myparagraph{Addressing the query agent} Commercial search engines often employ a query
intent identification method \cite{Jansen+al:07a} to improve reasoning
about the information need underlying a query. In a setting where a
query agent generates the query, agent (type) identification can
potentially also be of much merit. A case in point, suppose that the query
agent is based on a specific LLM. If the ranker agent is able to
identify which LLM that is, then a few opportunities emerge. To begin
with, automatic query reformulation can be performed using the exact same
LLM \cite{Jagerman+al:23a}. Such practice can potentially decrease the chances for query drift.
Identifying the query agent (type) can also help in selecting a ranking
function~\cite{Balasubramanian+Allan:10a}, especially if the document agent is also identified, as we further discuss below. Indeed, we show in Section \ref{sec:evaluation} that the alignment, or lack thereof, between the query, document and ranker agent can have considerable impact on search effectiveness.

\myparagraph{Addressing the document agent} Comparing document and query representations is a
fundamental relevance estimation paradigm. Document representation can
be induced using tf.idf, stochastic language models
\cite{Lafferty+Zhai:01a,Lavrenko+Croft:01a}, modern embedding
approaches \cite{e5,chen2024bge} and other approaches.  There has
also been a large body of work on enriching document representations
using cluster-based and topic-based information
\cite{Kurland+Lee:04a,Liu+Croft:04a,Wei+Croft:06a,Efron+al:12a}
and using queries automatically generated from a document
\cite{Nogueira+al:19a}.

In the multi-agent retrieval setting the ranker agent has to rank
documents, some of which were generated using document
agents. The potential ability to identify the document agent type can be of
much merit. Specifically, given our findings in Section \ref{sec:effectiveness-experiment}, and those in some recent work \cite{neural-llm-bias,dai2024cocktail}, alignment of the ranker agent and document agent can have significant impact on search effectiveness \cite{neural-llm-bias,dai2024cocktail}. Hence, an intriguing research direction is estimating document relevance based on the type of agent that generated the specific document.

Another direction worth exploration is the ``translation'' between
agent types. Such translation can be performed by the ranker agent
online during retrieval time or offline. Consider the following
example. Suppose that a query agent generates a query using some LLM
applied to a natural language description of an information need. On
the other hand, say that the document agent used a BERT-based
embedding approach~\cite{bert-transformer} to modify a human-authored document. (We devise and
evaluate such query and document agents in Sections \ref{sec:agents-implementation} and~\ref{sec:evaluation}.) We show in Section
\ref{sec:evaluation} that this misalignment has considerable impact on retrieval. Now, if the ranker agent knows how to ``translate''
the generated query, the document or both to a shared ground (e.g.,
LLM-based or embedding-based space), then presumably the retrieval would
be more effective. Conceptually similar considerations are common
in work on cross-lingual retrieval \cite{crossling}.

As noted above, in competitive search settings, document authors might be 
incentivized to modify their documents to improve their ranking for
queries of interest (i.e., search engine optimization). LLMs provide an effective grounds for such modifications, as recently shown \cite{Niv}. This reality brings about the need to devise ranker agents that account for the ranking incentives of document agents. For example, distilling content from the document which is due to pure authorship considerations rather than ranking incentives (e.g., keyword stuffing) is an important research direction (cf., \cite{Ziv-Ranker}). Kurland and Tennenholtz \cite{kurland_competitive_2022} present an elaborated discussion about the challenges of devising ranking functions in the face of ranking-incentivized document modifications.


\section{The Query Agent}
\label{sec:query-agent}
Query formulation is an opaque process in the lens of the search system: it only has access to the query with no insight into how it was constructed.
With the introduction of query agents, the interplay between user intent and the agent interpretation thereof also plays an important role.
This could introduce new complexities, biases, and challenges for the search engine~\cite{ranking-manipulation-conversational}.
Agents can generate queries that are of a different nature than those generated by humans~\cite{llm-query-variations}, and have different characteristics~\cite{query-generation-personality}. 
This necessitates developing query agents that formulate queries that reflect human intent while also accounting for potential biases. 

Query agents can be integrated with other agents. For example, an LLM-based conversational agent could invoke a lexical module to generate queries in order to mitigate the potential bias of ranker agents to certain query agent types. 

The ranker agent may operate with varying levels of knowledge about query generation~\cite{query-rewrite-for-rag}.
It could be aware of the query agent and the inputs (e.g., information need description if the query agent is part of the search system), it could have knowledge of the agent type (e.g., via API call signatures or query patterns), or it could remain entirely isolated from the generation process of the query. It is interesting to explore methods 
methods for making inference about the query agent under different knowledge states. 

\myparagraph{Decentralizing the ranker}
A query agent can search simultaneously with multiple rankers and/or multiple query variations and aggregate results based on different criteria.  
The merits of retrieving with more than one query were demonstrated~\cite{mult-queries-benefit}. 
This paradigm shift enables post-ranking interventions such as relevance feedback and fusion of ranked lists. 
Consequently, the ranking process becomes somewhat decentralized in the sense that the outputs of the ranker need not be the final output the user obtains. 

This creates opportunities for research on developing methods for the query agent to refine and aggregate search results through post-processing techniques.
Among the topics for future exploration are: (i) how to select ranker agents; (ii) how to formulate a query for each ranker agent; and (iii) how to fuse results retrieved by different rankers so as to satisfy the information need. Note that in contrast to standard fusion, which is often applied by meta-search engines that have no information about the underlying information need except for the query \cite{Kurland+Culpepper:18a}, here the query agent can know the information need. 

\myparagraph{Learning to Search} Unlike human users, who iteratively refine queries through manual reformulation, query agents can generate large-scale query variations and evaluate them in a systematic way. These agents can learn from past user interactions and generalize over users to refine query generation. 

\subsection{Strategic Query Agents}

\myparagraph{Matching the ranker agent} 
The query agent could employ different strategies. We show in Section~\ref{sec:effectiveness-experiment} that the alignment between the query and the ranker agents has a significant impact on retrieval effectiveness.
The query agent could employ methods to align with the ranker agent or a specific document agent. For example, when interacting with a lexical ranker agent, the query agent can estimate IDF values of terms~\cite{Yossef+Gurevich:08a} and formulate the query accordingly.

\myparagraph{Impact on publishers dynamics}
Both theoretical and empirical evidence suggests that the query agent has the potential to impact the dynamics between publishers~\cite{MultQueries}. Our analysis in Section~\ref{sec:online-experiment} further reinforces these findings. For example, prior work~\cite{MultQueries} demonstrated that an increased number of query variations makes equilibrium states in competitive retrieval unattainable.

\section{The Document Agent}
\label{sec:docAgent}

The document agent is responsible for generating or modifying documents. In the past, humans generated the documents, and agents' roles were limited to applying document modifications. However, LLMs are not only capable of performing document modifications but also of creating original content. Although the texts produced by LLMs may closely resemble human writing, there is still a ground difference in the way LLMs understand, interpret, and generate language~\cite{llm-parrots, how-llms-understand-language, tortured-phrases}.

\myparagraph{Non-Strategic Document Modifications}
In the competitive search ecosystem, document modifications are sometimes of strategic nature~\cite{Greg-bot, kurland_competitive_2022}.
Such modifications include, for example, keyword stuffing and link spamming~\cite{Gyongyi+Molina:05a}. 
These modifications are applied post-document generation and therefore require additional effort from the publisher. 
In contrast, LLM agents often facilitate the writing process.
Publishers therefore leverage these models for purposes beyond rank promotion, which can blure the distinction between strategic and non-strategic modifications~\cite{gpt-papers}.

How to differentiate between strategic and non-strategic modifications is already a challenging endeavor, even when considering only pure human-generated documents. Publishers could perform document modifications for a variety of reasons, such as reflecting temporal changes~\cite{temporal-web-dynamics}. The key distinction, however, lies in the impact on rankings: successful strategic modifications correlate with improved rankings, whereas non-strategic modifications need not necessarily result in improved ranking.
Future research should therefore go beyond classifying modifications as strategic or non-strategic. A possible direction could be to design mechanisms that actively de-incentivize strategic modifications without penalizing non-strategic updates. 

\subsection{Strategic Document Manipulations}

\myparagraph{Mimicking the Winner}
A well-documented phenomenon in competitive search environments is the tendency of less successful publishers to mimic the strategies of top-ranking competitors~\cite{Nimrod,Greg-bot,Greg-Herding,MultQueries,Lemss}.
This behavior, often referred to as "mimicking the winner," has been observed in ranking competitions among humans~\cite{Nimrod}, ranking competitions of humans with document agents~\cite{Greg-bot,Niv,MultQueries} and between document agents themselves~\cite{Lemss}. \citet{Nimrod} demonstrated that when publishers observe a dominant strategy leading to high rankings, they are incentivized to adopt similar modifications. This behavior leads to homogenization of documents and reduction in the diversity of the corpus~\cite{Greg-Herding}.

\myparagraph{Adapting to Ranker Biases}
It was recently shown \cite{neural-llm-bias, dai2024cocktail} that LLM rankers are biased towards LLM-generated documents.
In a competitive environment, publishers may find themselves in a challenging position where human-generated content could be deprioritized in favor of LLM-generated alternatives.

Suppose that this very paper were to be submitted for review at a conference.
If we suspect that LLMs influence the review process~\cite{conference-llm-reviews}, whether
through automated initial assessments or reviewers using LLM
tools for comprehension, then, as authors, we are incentivized to
optimize our writing for these models. This creates a paradox: the
very tools designed to enhance evaluation may instead undermine its integrity by encouraging content to cater to algorithmic biases rather than relevance.

\myparagraph{From query attacks to query-agent attacks}
Many existing attacks on ranking functions, in the form of document modifiications, are query-specific~\cite{Wang+al:22a, Greg-Herding,Liu+al:22a,Song+al:22a,Wang+al:22a,Wu+al:22a,Chen+al:23a,Chen+al:23b}. By generating multiple queries and performing aggregation, etc., query agents could perhaps be able to generate queries that can mitigate the impact of these specific attacks. 
Hence, attack strategies of document agents may shift toward targeting query agents~\cite{GEO}.
This further strengthens our call for future research that explores methods to mitigate the bias of LLM rankers in favor of LLM-generated content, as discussed in Section~\ref{sec:devising-ranker-agents}.
The study of such potential attacks and the means to counteract them is an important research direction.

\section{Re-Considering Evaluation for Multi-Agent Retrieval Settings}
\label{sec:evalConsider}
There is a long history of work on evaluation of ad hoc retrieval methods. Perhaps the most fundamental paradigm is Cranfield's \cite{cranfield}: using a test collection with documents, queries and relevance judgments. The Cranfield paradigm was the basis for foundational evaluation initiatives: TREC, CLEF, NTCIR, INEX, etc. 

The question of whether retrieval performance patterns observed for one test collection will transfer to others is of utmost importance. For example, in the relatively early days of Web search, studies showed that classical retrieval methods which were highly effective on TREC's newswire collections failed over Web collections \cite{Singhal+Kaszkiel:01a}. This demonstrates, among others, the effect of document agents --- humans, back in these days. From a user perspective, recent work shows that the relative performance patterns of ranking functions can significantly depend on the quality of the query selected to represent the information need \cite{shane:22a,Rashidi+al:24a}. 

In Section \ref{sec:evaluation} we demonstrate the substantial effect of the interplay between ranker, document, and query agents on retrieval performance. A case in point, using for evaluation a corpus composed only of human-generated documents can lead to completely different findings about the relative effectiveness of ranker agents with respect to a corpus composed of both human and LLM-based generated documents. We also show that the alignment, or lack thereof, of the types of query and document (or ranker) agents has considerable impact. At the same time, creating document and query agents (e.g., using different LLM prompts) is quite an easy task. 

The reality just described is a call for arms to re-consider evaluation in multi-agent retrieval settings. The main question is how should test collections be constructed in terms of agents generating documents and queries, and to what extent can findings for one collection transfer to another given the extreme variability and fast-paced changes in agent technology. 

Another important aspect is evaluation in competitive search settings \cite{kurland_competitive_2022} where document authors modify their documents so as to improve their future ranking (search engine optimization). As noted above, there is an increasing body of work on document agents that modify documents for rank promotion while preserving content quality \cite{Greg-bot,Niv}. Static collections cannot be used to evaluate new retrieval methods, for example, as document agents respond to specific rankings. This state-of-affairs calls for simulation-based evaluation in multi-agent retrieval settings (cf., \cite{Lemss,kurland_competitive_2022}) where designers of new retrieval methods (ranker agents) can evaluate them in an online manner. While there has been work on simulating users for interactive retrieval \cite{Balog+Zhai:24a}, integrating document and ranker agents is an important venue for future work.

\section{Empirical Exploration}
\label{sec:experimental-settings}

The main goal of the experiments presented below is to demonstrate the effects on the retrieval setting of interactions between different query, document, and ranker agents.
To this end, we conduct a series of three experiments. 
In the first experiment, we evaluate the effectiveness of different rankers
with various query and document agents (Section~\ref{sec:effectiveness-experiment}).
In the second experiment, we explore the interplay between the document and ranker agent, specifically when document agents compete against human-authored documents for rank promotion (Section~\ref{sec:offline-experiment}).
In the third experiment, we
study how ranker and query agents influence the competitive dynamics among different document agents (Section~\ref{sec:online-experiment}).

We now turn to describe our methodology. First, we detail the datasets used for evaluation (Section~\ref{sec:data}). Then, in Section~\ref{sec:agents-implementation}, we introduce the implementations of the ranker, query, and document agents, each categorized into three types: lexical, semantic, and LLM-based. Additionally, for query and document agents, we also consider human agents, i.e., queries and documents generated by humans.
Several of the agents we present are novel to this study.

\subsection{Data}
\label{sec:data}

We utilize datasets from ranking competitions in which human participants competed against automated agents ~\cite{Greg-bot, Niv, MultQueries}. These competitions were structured to simulate realistic retrieval environments where documents evolved across multiple ranking rounds due to ranking incentives. The datasets consist of documents that were iteratively modified by human participants attempting to improve their ranking, alongside automated agents that applied ranking-incentivized modifications.

We use four publicly available datasets~\cite{Greg-bot, Niv, MultQueries} with recordings of content-based ranking competitions held in IR courses where students acted as publishers. The students were given an initial document and were tasked with promoting it in rankings induced by an undisclosed ranking function over several rounds. At the
beginning of the competition, all students received the same example of a relevant document for each topic. After each round, they were presented with a ranking of their documents.
Their objective was to modify their documents in an effort to improve their ranking in the subsequent round. All documents were plaintext and limited to a maximum of $150$ terms. The competitions were incentivized through performance-based course-grade bonuses and were conducted with the approval of ethics committees.

\begin{table}
\caption{Datasets summary.}
    \centering
    \scriptsize
    \begin{tabular}{c|c|c|c|c}
        Data &  \GregDataset~\cite{Greg-bot}& \NivDataset~\cite{Niv} &\MultiB~\cite{MultQueries} & \MultiD~\cite{MultQueries}\\
        \hline
        \# Docs & 300 & 600 & 1500& 1500 \\
        \# Human Docs & 120 & 120 & 630 & 610 \\
        \# Rounds & 4 & 4 & 10 & 10 \\
        \# Queries & 15 & $15 \times 2$& 30 & 30 \\
    \end{tabular}
    \label{tab:dataset}
\end{table}

The competitions conducted by~\citet{Greg-bot} employed a \lambdamart
ranker \cite{wu2010adapting} with $25$ features: $24$ lexical
features, and a single document quality score.  The competitions were held for
$15$ queries (topic titles) from TREC's ClueWeb09 dataset.

The competitions held by~\citet{Niv} employed the \efive-based ~\cite{e5} ranking function\footnote{intfloat/e5-large-unsupervised}. Two competitions were held for each of the $15$ same queries as Goren et al.'s~\cite{Greg-bot} competition.

As discussed in Section~\ref{sec:query-agent}, the query agent might consider different query variations, generated by either one or a variety of query agent types. 
To explore the effects of query variations on the competition dynamics, we utilize the dataset from \citet{MultQueries}. In these competitions students were requested to improve their rankings for three different query variations per topic. We use two of their competitions: the ones where students were not allowed to use \llm tools.
In the first competition, the ranker was a \bert~\cite{BERT} model fine-tuned on the MS-MARCO dataset.
The second competition employed LambdaMART trained specifically for competitive setings~\cite{Ziv-Ranker}.
The features were all lexical except for one semantic feature based on the score of the same \bert model used in the first competition.  
Both competitions were conducted for $30$ topics from TREC's ClueWeb12 collection, with query variations from the dataset reported in~\cite{uqv100}.

We use \firstmention{\single} to refer to competitions where students competed for a single query and \firstmention{\multi} 
to refer to competitions with multiple query variations. Similarly, \firstmention{\ltr} denotes competitions using a LambdaMART,
while \firstmention{\neu} refers to those using a neural ranker (\bert / \efive). 

All competitions featured a total of five participants per topic. However, not all of them were students, as some of the competing documents were planted. 
In addition, each competition included different \botagents that competed alongside human participants. The students were unaware that they were competing against automated agents. The \botagent in \GregDataset utilized both lexical and semantic features, whereas in the other competitions, the \botagent was entirely \llm-based.
Table~\ref{tab:dataset} summarizes the details of all four datasets used in our experiments.

\subsection{Agents Implementation}
\label{sec:agents-implementation}
\myparagraph{Ranker}
We use
three types of agents (lexical, semantic and LLM) each with two rankers.
\subsubsection*{\lexicalagent}
We use \bm and \tfidf\footnote{The retrieval score is the sum of \tfidf weights of query terms in the document.}  with Indri's default hyperparameter values\footnote{\url{https://www.lemurproject.org/indri}}.
Collection statistics are based on ClueWeb09 for \GregDataset and \NivDataset and ClueWeb12 for \MultiB and \MultiD.

\subsubsection*{\semanticagent}
We use \contriever~\cite{contriever} and \efive~\cite{e5}, both fine-tuned on the MS-MARCO~\cite{msmarco} dataset\footnote{intfloat/e5-base-v2, nthakur/contriever-base-msmarco}. The document retrieval score is the cosine between the query and document embedding vectors.

\subsubsection*{LLM} 
We adopt a pointwise relevance generation model~\cite{liang2022holistic} for an LLM ranker.
We experiment with two lightweight ($<10$B parameters) instruction-tuned open-source LLMs: \llamaWithVersion~\cite{dubey_llama_2024} and \gemmaWithVersion~\cite{gemma_team_gemma_nodate} from the Hugging Face repository\footnote{meta-llama/Meta-Llama-3.1-8B-Instruct, google/gemma-2-9b-it}. 

All the pre-trained models described above are used consistently across all experiments with their Hugging Face default hyperparameter values, unless specified otherwise.

\myparagraph{Query}
To
study the impact of different query formulation techniques, we implemented several query agents that generate queries based on the backstories provided in the UQV100 dataset~\cite{uqv100}.

Herein, we treat each query variation independently as a query, i.e., without directly considering the combined effect of multiple queries simultaneously generated by the query agent.
Previous work~\cite{MultQueries} explored the dynamics when humans and document agents compete for multiple (human) queries simultaneously. 
The exploration of the dynamics when document agents compete for multiple queries generated by different query agents is beyond the scope of this paper and remains an interesting direction for future work.

\subsubsection*{Human} 
We utilize the human-generated query variations from the UQV100 dataset, which was created via crowdsourcing~\cite{uqv100}. For each topic, we selected the five most frequent query variations. 

\subsubsection*{Lexical}
We applied the YAKE keyword extraction method~\cite{YAKE} to identify key phrases in the backstories. Each extracted phrase was considered a potential query and was ranked w.r.t. the backstory using BM25. The five highest-scoring phrases per topic were selected.

\subsubsection*{Semantic} 
We adopt the doc2query method~\cite{doc2query} to generate for each topic a pool of $1,000$ query variations from the backstory.
We use two representations: \efive and \contriever.
For each of them, the final five variations for each topic were selected based on the cosine similarity of the embedding vector of the variation and that of the backstory.

\subsubsection*{LLM}
We utilized an existing dataset containing query variations generated by \gptWithVersion~\cite{llm-query-variations} and replicated its methodology to generate additional variations with \llama~\cite{dubey_llama_2024} and \gemma~\cite{gemma_team_gemma_nodate}, all set with a temperature of $1$.
To ensure well-formed outputs, we added to the prompt an instruction to return plain lists of queries. For each topic, the first five variations generated by each LLM were selected.

\myparagraph{Document}
We implement different document agents that apply ranking-incentivized modifications to documents. 
The \botagents operate as follows: first, they observe the ranking for a given query. Then, they apply modifications to a document to improve its ranking in the next round.

Two
ranking-incentivized \botagents for document modifications were recently presented.
~Gorent et al.'s \cite{Greg-bot} \botagent replaces a sentence \SrcSentence from the
document with a candidate sentence \TargetSentence. 
The \botagent selects the pair $(\SrcSentenceMath, \TargetSentenceMath)$ using a learning-to-rank approach with a small set of lexical and semantic features. 
Our goal is to compare \textit{lexical} and \textit{semantic} agents, making it unsuitable to adopt their method directly.
Instead, we develop a \botagent inspired by their approach.
Bardas et al. ~\cite{Niv} used LLM-based agents for document modification. We adopt their approach to develop our LLM \botagent.

\subsubsection*{Lexical}

Inspired by~\citet{Greg-bot}, the lexical \botagent modifies documents
by replacing a sentence from the original document, \SrcSentence (source), with
a candidate sentence \TargetSentence (target).  Candidate sentences are
extracted from documents published by other publishers in the round.
The \botagent developed by \citet{Greg-bot} relies on a set of lexical and semantic features.  In particular, it employs two lexical
features, assuming that higher feature values indicate increased
retrieval score assigned by the undisclosed ranking function:
\firstmention{\QryTerm}, which is the fraction of query term occurrences in the sentence, and \firstmention{\SimTop}, which represents the cosine similarity between the sentence's \tfidf vector and the centroid of the top \numhighestranked ranked documents in the current ranking.

Each feature is computed for both the source (\QryTermSrc, \SimSrcTop) and target (\QryTermTarget, \SimTargetTop) sentences. The final score is:
\begin{equation*}
    score_{lex}(\SrcSentenceMath, \TargetSentenceMath) = \interpolationMath \cdot (\QryTermTargetMath - \QryTermSrcMath) + (1-\interpolationMath) \cdot (\SimTargetTopMath-\SimSrcTopMath).
\end{equation*}
The highest scoring sentence pair is selected to perform the replacement.
To ensure that modifications do not significantly alter
the document, we only consider a pair $(\SrcSentenceMath, \TargetSentenceMath)$ if the \tfidf-based cosine similarity between \SrcSentence and \TargetSentence exceeds a predefined threshold \nlithreshold.
The parameter values are: $\interpolationMath \in [0, 0.1, \ldots,1]$, $\numhighestrankedMath \in \{2, 3, 4\}$ and $\nlithresholdMath \in [0,0.1,\ldots,0,5]$.

\subsubsection*{Semantic}
The semantic document agent, novel to this study, modifies documents in a similar manner to the lexical document agent, but relies on semantic (embedding) representations rather than lexical features. 
We
use three strategies to select the candidate target sentences to use
for replacement in the original document.
The first strategy, \firstmention{\all}, considers sentences from all
other documents as candidates.  The \firstmention{\better} strategy
selects only sentences from documents ranked higher than the original
document.  The \firstmention{\best} strategy picks sentences
exclusively from the highest-ranked document.  For each pair of
sentences $(\SrcSentenceMath, \TargetSentenceMath)$ and query $q$, we
compute the cosine similarity between the embedding vectors
representing \SrcSentence and \TargetSentence and the similarity between \TargetSentence and $q$: 
\begin{equation*}
    score_{sem}(\SrcSentenceMath, \TargetSentenceMath) = \interpolationMath \cdot cos(\SrcSentenceMath, \TargetSentenceMath) + (1-\interpolationMath) \cdot cos(\SrcSentenceMath, q)
\end{equation*}
We use this method to balance the presumed relevance of the candidate (target)
sentences, and accordingly their contribution to the retrieval score, and their
contextual fit within the document.  To preserve document coherence,
we use an NLI model~\cite{NLI} to ensure that the new sentence is entailed
by the candidate sentence. A candidate sentence is considered only if
it entails the original sentence, which is considered to be the case
if the probability assigned by the NLI model is higher
than \firstmention{\nlithreshold}.  In our experiments, we use two
dense pretrained text embedding models before
fine-tuning: \efive~\cite{e5}
and \contriever~\cite{contriever}\footnote{intfloat/e5-base-unsupervised,
facebook/contriever}.
The value of \interpolation was selected from $[0, 0.1,\ldots,1]$ and the value of \nlithreshold was selected from $[0,0.1,\ldots,0.5]$. 

\subsubsection*{LLM}

The \textit{LLM-based} \botagent introduced by \citet{Niv} modifies documents using different prompting strategies.
Document agents are defined by a specific LLM model and the prompting strategy.
The prompts include the rules and constraints for the document format, the original document, and the assigned query.
Additionally, they include information about past documents and their rankings. 
We use two prompting strategies: \feedbackPair, which includes two randomly selected documents and their rankings, and \feedbackAll, which provides the full ranked list. 
To mitigate copying behavior observed in human-driven ranking competitions~\cite{Nimrod, Greg-Herding}, the \nocopy\xspace flag ensures that the \botagent is also explicitly warned against directly copying content from other documents. 
Both \llama and \gemma were used with $top_p=0.95$ and $temperature=0.7$.

Unless otherwise specified, the free-parameter values of all document agents were set per ranker to maximize \firstmention{scaled rank promotion}: the raw rank change of a document, normalized by the maximum possible promotion or demotion based on its position. Higher values indicate increased improvement.

\section{Results and Analysis}
\label{sec:evaluation}
\subsection{Effectiveness of Rankers with Different Query Agents}
\label{sec:effectiveness-experiment}

\subsubsection{Setup}
In this experiment, we assess the effectiveness of different ranker agents when queries and documents are generated by
query and document agents, respectively. 
A \firstmention{\agentcorpus} is a set of documents that are generated by a collective of document agents.
We use three
corpora: (i) \firstmention{\purehumancorpus}, which contains only human-generated documents; (ii) \firstmention{\purellmcorpus}, which consists solely of documents generated by the \llm document agent; and (iii) \firstmention{\mixedcorpus}, which includes a combination of both human and \llm-generated documents. 
The query agents generate query variations based on the backstory corresponding to each topic. We consider both query variations generated by humans and variations of queries generated by different types of query agents.

For evaluation, we use the two \multi datasets, \MultiB and \MultiD,  with competitions for multiple queries.
These are the only datasets with a corresponding backstory, which is required by the query agent.
We evaluate the effectiveness of ranker agents on rounds $2$-$10$, since the \llmagent documents were not modified in the first round. 

A total of five documents are ranked for each query, with $2$-$3$ authored by humans and the remaining generated by the LLM agent. 
Hence, we measure the effectiveness of different rankers using the nDCG@1 metric. Statistical significance is determined based on a two-tailed paired t-test with $p < 0.05$ and Bonferroni correction for multiple comparisons.

We address three research questions:
\textbf{RQ1}: How does retrieval effectiveness change when rankers operate on corpora composed of documents created by different document agents? 
\textbf{RQ2}: Do ranker agents perform differently depending on the query agent?
\textbf{RQ3}: How is ranker agent effectiveness
affected by the document and query agent types (i.e., human, lexical, semantic and LLM)?

\subsubsection{Evaluation}

A total of five documents are ranked for each query, with $2$-$3$ authored by humans and the remaining generated by the LLM agent. 
Hence, we measure the effectiveness of different rankers using the nDCG@1 metric. 
The NDCG values are quite high, as in past work~\cite{Ziv-Ranker} as most documents in the competitions were relevant~\cite{Nimrod, MultQueries}.
Statistical significance is determined based on a two-tailed paired t-test with $p < 0.05$ and Bonferroni correction for multiple comparisons.

\subsubsection{Results}

\begin{table}
    \caption{Mean nDCG@1 (scaled by 100) for different ranker and query agents averaged over queries and ranker agents of the same type. The differences between the NDCG@1 values of human and LLM corpora for human queries, as well as the differences between \human or \llm corpus and the \mixed corpus, are all statistically significant.}
    \centering
    \scriptsize
    \begin{tabular}{llcc}
        \toprule
        & & \multicolumn{2}{c}{\textbf{query agent}} \\
        \cmidrule(lr){3-4}
        \textbf{ranker agent} & \textbf{corpus} & \textbf{human} & \textbf{LLM (\gpt)} \\
        \midrule
        \multirow{3}{*}{lexical} 
        & \human  & 92.639 & 92.701 \\
        & \llm    & 91.802 & 92.108 \\
        & \mixed  & 90.148 & 90.056 \\
        \midrule
        \multirow{3}{*}{semantic} 
        & \human  & 92.340 & 92.176 \\
        & \llm    & 91.466 & 92.049 \\
        & \mixed  & 89.185 & 89.877 \\
        \midrule
        \multirow{3}{*}{llm} 
        & \human  & 92.781 & 92.148 \\
        & \llm    & 91.327 & 91.954 \\
        & \mixed  & 89.821 & 90.009 \\
        \bottomrule
    \end{tabular}

    \label{tab:ndcg1_results}
\end{table}

Table~\ref{tab:ndcg1_results} reports for each agent type the average nDCG@1 over the two ranker agents of the same type. 
We see that \highlight{across all rankers and query agent types,  retrieval performance is significantly lower on a mixed corpus (i.e., corpus composed of documents created by human and LLM agents)
compared to a corpus with documents generated by a single agent (human or LLM)} (\textbf{RQ1}). For human-generated queries, the rankers consistently performed better on a \purehumancorpus corpus than on a \purellmcorpus corpus. However, for LLM-generated queries, performance was similar across the \purehumancorpus and \purellmcorpus corpora.

\begin{table}
    \caption{Mean nDCG@1 (scaled by 100) using the \textbf{\contriever} ranker across different query agents and corpora. '$c$' ('$h$') marks a statistically significant difference with \contriever (human)-generated queries for the same corpus.
    }
    \centering
    \scriptsize
    \begin{tabular}{l|c|c|ccc}
        \toprule
        & \multicolumn{5}{c}{\textbf{Query Agent}} \\
        \cmidrule(lr){2-6}
        \textbf{Corpus} & \textbf{Human} &\textbf{\contriever} & \textbf{\gpt} & \textbf{\gemma} & \textbf{\llama} \\
        \midrule
        \human  & $91.747$ & $92.167$ & $91.438$ & $90.981^{hc}$ & $90.531^{hc}$ \\
        \llm    & $91.160$ & $91.006$ & $91.562$ & $90.821$ & $91.352$ \\
        \mixed  & $88.568^{c}$ & $89.815^{h}$ & $88.802^{c}$ & $88.253^{c}$ & $88.179^{c}$ \\
        \bottomrule
    \end{tabular}
    \label{tab:ndcg1_contriever}
\end{table}

The results in Table~\ref{tab:ndcg1_contriever} reveal that retrieval effectiveness varies not only between human and LLM-generated queries but also significantly among different query agents of the same type. 
For example, Table~\ref{tab:ndcg1_contriever} shows that the \contriever ranker agent achieves $NDCG@1$ of $92.167$ when queries are generated by a \contriever query agent on the \purehumancorpus corpus, but it drops to $90.531$ when the query agent is \llama. 
However, it drops only to $91.438$ when the queries are generated by a different LLM: \gpt. 
The results in Table~\ref{tab:ndcg1_contriever} demonstrate that ranker agents exhibit not only a difference between human and LLM-generated queries (that is, different query agent types), but also substantial variance across different LLMs (i.e., different query agents of the same type) (\textbf{RQ2}).
This suggests that ranking agents are not uniformly effective across different query agents and document agents. Moreover, we see in Table~\ref{tab:ndcg1_contriever} that the performance of the ranker agent is quite the same for different query agents when the documents are all generated by the same \llm document agent (\textbf{RQ3}).
Table~\ref{tab:ndcg1_contriever} also shows that the highest effectiveness for both human and mixed corpora for the \contriever ranker agent is when the queries are generated by the matching query agent (\contriever), which attests to the importance of alignment between the query agent and the ranker agent.

The findings presented above underscore the
importance of using corpora with documents generated by both LLMs and
humans for evaluation. Corpora composed only of human-authored
documents do not represent the real world anymore, and the resultant
evaluation can be biased. Moreover, evaluation should also be based on a variety of query agents.

\subsection{Offline Evaluation of Document Agents}
\label{sec:offline-experiment}
\subsubsection{Setup}
We perform the offline evaluation with the two \single\xspace datasets: \GregDataset and \NivDataset. We cannot use the \multi datasets since our goal here is to contrast the effectiveness of document agents with respect to humans in promoting documents in rankings. The \botagents are designed to promote the document for a single query, while the students in the \multi competitions modified their documents to promote them for multiple queries. 

The goal of this experiment is not to develop the most effective \botagent but rather to examine how different document agents interact with ranker agents. 
Our research questions are:
\textbf{RQ4}: Can the zero-shot lexical and semantic document agents we developed successfully compete for rankings against human publishers?
\textbf{RQ5}: Does the misalignment between the \botagent and the ranker agent types affect the \botagents' effectiveness in promoting their documents? 
\textbf{RQ6}: How well can document agents perform when knowing only the ranker agent type (i.e., lexical, semantic, or \llm)?

We assess the \botagents' ranking-promotion using the {scaled rank promotion} metric described in Section \ref{sec:agents-implementation}.

\subsubsection{Results}

\begin{table}
  \caption{The scaled rank promotion of the
    \botagents, averaged across the \GregDataset and \NivDataset datasets.
    Boldface: the highest promotion in a column.}
    \centering
    \scriptsize
    \begin{tabular}{l|l|cc|cc|cc}
        \toprule
        \multicolumn{2}{c}{ } &\multicolumn{6}{|c}{\textbf{Ranker Agent}} \\
        \cmidrule(lr){3-8}
        \multicolumn{2}{c|}{\textbf{Document Agent}} 
        & \multicolumn{2}{c|}{\textbf{Lexical}} 
        & \multicolumn{2}{c|}{\textbf{Semantic}} 
        & \multicolumn{2}{c}{\textbf{LLM}} \\
        \hline
        \textbf{Type} & \textbf{Model} 
        & \textbf{\bm} & \textbf{\tfidf} 
        & \textbf{\contriever} & \textbf{\efive} 
        & \textbf{\gemma} & \textbf{\llama} \\
        \hline
        Human    & -         & 0.071  & 0.074  & 0.179  & 0.183  & 0.117  & 0.078  \\
        Lexical   & -         & \textbf{0.430}  & \textbf{0.433}  & 0.253  & 0.236  & 0.158  & 0.049  \\
        Semantic  & \contriever & 0.269  & 0.267  & \textbf{0.363}  & 0.289  & 0.226  & 0.094  \\
        Semantic  & \efive   & 0.229  & 0.233  & 0.308  & \textbf{0.329}  & 0.183  & 0.092  \\
        LLM       & \gemma     & 0.217  & 0.216  & 0.273  & 0.264  & \textbf{0.497}  & 0.280  \\
        LLM       & \llama     & 0.084  & 0.089  & 0.268  & 0.240  & 0.354  & \textbf{0.640}  \\
        \bottomrule
    \end{tabular}
    \label{tab:bot_performance_large}
    \vspace{-4mm}
\end{table}

Table~\ref{tab:bot_performance_large} presents the average scaled rank promotion of the different \botagents. For each ranker agent and \botagent, we present the average scaled promotion of the most effective document agent, i.e., the one whose hyperparameter values were set to optimize scaled rank promotion. Table~\ref{tab:bot_performance_large} shows that the average scaled promotion of all \botagents is positive, which means that, on average, \botagents are always successful in promoting their documents compared to humans, regardless of their type and the ranker agent's type. Specifically, the novel zero-shot lexical and semantic \botagents we developed consistently achieve higher scaled rank promotion values than humans across all ranker agents (\textbf{RQ4}). 
The same finding holds for the LLM document agents as recently shown \cite{Niv}.
This shows that all rankers are vulnerable to ranking-incentivized document manipulations by document agents. 

Table~\ref{tab:bot_performance_large} shows an additional clear trend: \highlight{A mismatch between the document agent type and the ranker agent type leads to a substantial decrease in the ability of the \botagent to improve the ranking of its document}. 
Conversly, when the document agent and ranker agent types are aligned, 
the document agent's ability to achieve higher rankings improves markedly (\textbf{RQ5}).
For example, the \gemma \botagent achieved a scaled rank promotion of $0.497$ when the ranker agent was also \gemma, but only $0.354$ when the ranker agent was \llama. When the ranker agent type was of a different type (i.e., not an LLM), the maximal scaled promotion achieved was only $0.226$.
Similar trends are observed for the other document agents. 
Rank promotion is highest when the document agent and the ranker agent are the same.  
The second highest rank promotion is attained when the ranker agent is of the same type as the document agent.
This indicates that the alignment between the ranker agent and the document agent is crucial from the publishers' perspective for rank promotion strategies.
Moreover, the findings suggest that the publisher can be highly effective even with knowledge of only the ranker agent type (i.e., lexical, semantic, or \llm) (\textbf{RQ7}).

\subsection{Online Simulation: Long-Term Multi-Agent Interactions}
\label{sec:online-experiment}
\subsubsection{Setup}

Our online evaluation consists of a simulation of a ranking competition between different \botagents. 
We use the documents from the last round of the \MultiB and \MultiD datasets as a starting point for the simulation. 
The simulation was performed using the CSP framework~\cite{Lemss}. 
We employ two semantic \botagents: \efive and \contriever with the candidate sentence selection strategy \textit{\better}, 
$\nlithresholdMath=0.5$ and $\interpolationMath=0$.
Two \llm \botagents are also included in the simulation: \gemma and \llama, both with \feedbackPair \xspace and \withoutnocopy. (See Section \ref{sec:agents-implementation} for details about the 
agents\footnote{Hyperparameter values were set to maximize rank promotion performance w.r.t. the ranker agent for which the promotion was minimal in the offline evaluation.}.)
Additionally, we use as a reference comparison a static agent that does not modify her initial document throughout the rounds. 
The agents are randomly paired with an initial document from the last round of the competition.
We ran the simulation for four rounds for each of the query variations, which results in $30$ 
different 
competitions for each set of documents.  
In total, we ran
$1800$ simulations. 
Our research question is \textbf{RQ7:} How do different query agent types impact the dynamics between different document agents that compete against each other for document rankings? 
The effectiveness of \botagents is measured by the average rank position of their documents. 



\subsubsection{Results}

\begin{table}
    \caption{The average rank position (over all competitions and rounds) of the static document agent's document.}
    \centering
    \scriptsize
    \begin{tabular}{l|ccc}
        \toprule
        & \multicolumn{3}{c}{\textbf{Ranker Agent}} \\
        \cmidrule(lr){2-4}
        \textbf{Query Agent} & \textbf{\lexicalagent} & \textbf{\semanticagent} & \textbf{\llm} \\
        \midrule
        human    & 2.79 & 3.09 & 3.19 \\
        lexical  & 3.26 & 3.45 & 3.32 \\
        semantic & 3.10 & 3.33 & 3.17 \\
        llm      & 2.99 & 3.19 & 3.26 \\
        \bottomrule
    \end{tabular}
    \label{tab:static_player}
\end{table} 

We begin by evaluating the performance of the static document agent. The variance of its performance across different query agents indicates fundamental differences in the settings' dynamics. We see in Table~\ref{tab:static_player} that the static \botagent demonstrates superior performance when the query agent type and ranker agent type are not aligned.
For example, the average rank of the static \botagent is $3.26$ when both the query agent and the ranker agent are lexical, but the average rank is $2.99$ when the query agent type is LLM. 
This means that when the \highlight{query and ranker types are mismatched, the ability of publishers to perform ranking-incentivized document manipulations decreases.}  This finding is in line with that presented above about the impact of the alignment between the document agent and the ranker agent type on the document agent rank-promotion.

\begin{figure}
    \centering
    \scriptsize
    \begin{subfigure}{0.3\textwidth}
        \centering
\includegraphics[width=\linewidth]{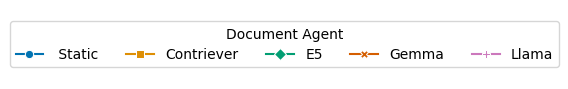}
    \end{subfigure}
    \begin{subfigure}{0.45\textwidth}
        \centering
        \includegraphics[width=0.9\linewidth]{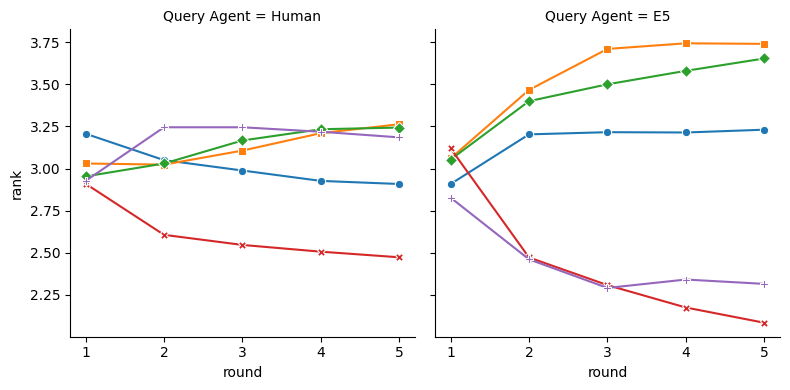}
        \caption{Ranker Agent: \efive.}
        \label{fig:e5_misalignment}
    \end{subfigure}
    \hfill
    \begin{subfigure}{0.45\textwidth}
        \centering
        \includegraphics[width=0.9\linewidth]{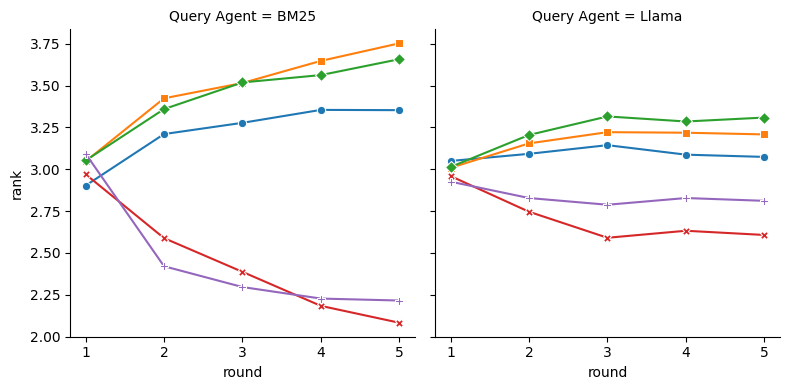}
        \caption{Ranker Agent: \bm.}
        \label{fig:bm25_misalignment}
    \end{subfigure}
    \caption{The average rank of document agents in each round across different ranker agents and query agents.}
\label{fig:query_ranker_misalignment}
\end{figure}

Diving deeper into the missaligment between the ranker agent type and the query agent type, we see in Figure~\ref{fig:query_ranker_misalignment} that the performance of \botagents is highly affected by the query agent type and its alignment with the ranker agent type. 
When the ranker agent and query agent types are aligned, the difference between different \botagents is higher (recall that this is a zero-sum game; the success of one \botagent comes at the expense of the other). 
In Figure~\ref{fig:e5_misalignment} we also observe the resultant rank sensitivity of certain document agents to the query agent. For example, when queries were generated by \efive, the \llama agent was the second-best \botagent. However, when queries were generated by humans, its performance declined and resulted in a substantially lower average rank.

\paragraph{Implications for the Ranker Agent}
We already observed in Section~\ref{sec:effectiveness-experiment} that when the query and document agent types are mismatched, the effectiveness of the ranker diminishes. 
However, an open question remains: how does this misalignment specifically impact modification strategies and the performance of \botagents in competitive settings? Understanding this effect is important for designing ranker agents that account for adversarial adaptation and strategic dynamics.

\paragraph{Implications for the Document Agent}
The results of the offline experiment (Section~\ref{sec:offline-experiment}) suggest that the alignment between the document agent and the ranker agent has high influence on the ability of the publisher to promote her document. Here, the results show that the alignment between the document and the ranker agent is not the only factor that affects the document agent's performance. 
We also showed that the dynamics are highly affected by the query agent. 
An interesting direction for future research is designing strategic query agents that promote desired dynamic outcomes in terms of search effectiveness and corpus effects.

\section{Conclusions}
The progress with large language models (LLMs) has opened the era of agents, specifically, in the information retrieval (IR) field \cite{Chen+al:21a,Shah+White:24a,White:25a,Zhai:24a,Zhang+al:24a}. We argued that the most fundamental task in the information retrieval field, namely, ad hoc retrieval, should now be re-visited in the face of multi-agent retrieval settings.

Document authors use agents to generate content. Users of search engines use agents as assistants for their information seeking tasks. There are novel opportunities for agent-based retrieval models. We argued that existing retrieval frameworks and paradigms, as well as evaluation, should be re-considered in the agent era. We also discussed fundamental research directions we believe are important for the study of multi-agent retrieval settings. In addition, we empirically analyzed an elaborated suite of multi-agent retrieval settings. Our findings shed light on the far reaching implications of the inter-play between document, user and ranker agents.

\newpage
\balance
\bibliographystyle{ACM-Reference-Format}
\bibliography{bib,orenBib, cunlp-ir}


\begin{thebibliography}{78}


\ifx \showCODEN    \undefined \def \showCODEN     #1{\unskip}     \fi
\ifx \showDOI      \undefined \def \showDOI       #1{#1}\fi
\ifx \showISBNx    \undefined \def \showISBNx     #1{\unskip}     \fi
\ifx \showISBNxiii \undefined \def \showISBNxiii  #1{\unskip}     \fi
\ifx \showISSN     \undefined \def \showISSN      #1{\unskip}     \fi
\ifx \showLCCN     \undefined \def \showLCCN      #1{\unskip}     \fi
\ifx \shownote     \undefined \def \shownote      #1{#1}          \fi
\ifx \showarticletitle \undefined \def \showarticletitle #1{#1}   \fi
\ifx \showURL      \undefined \def \showURL       {\relax}        \fi
\providecommand\bibfield[2]{#2}
\providecommand\bibinfo[2]{#2}
\providecommand\natexlab[1]{#1}
\providecommand\showeprint[2][]{arXiv:#2}

\bibitem[Abbasiantaeb et~al\mbox{.}(2024)]%
        {mult-queries-benefit}
\bibfield{author}{\bibinfo{person}{Zahra Abbasiantaeb}, \bibinfo{person}{Simon Lupart}, {and} \bibinfo{person}{Mohammad Aliannejadi}.} \bibinfo{year}{2024}\natexlab{}.
\newblock \bibinfo{title}{Generating Multi-Aspect Queries for Conversational Search}.
\newblock
\newblock
\showeprint[arxiv]{2403.19302}~[cs.IR]
\urldef\tempurl%
\url{https://arxiv.org/abs/2403.19302}
\showURL{%
\tempurl}


\bibitem[Aggarwal et~al\mbox{.}(2024)]%
        {GEO}
\bibfield{author}{\bibinfo{person}{Pranjal Aggarwal}, \bibinfo{person}{Vishvak Murahari}, \bibinfo{person}{Tanmay Rajpurohit}, \bibinfo{person}{Ashwin Kalyan}, \bibinfo{person}{Karthik Narasimhan}, {and} \bibinfo{person}{Ameet Deshpande}.} \bibinfo{year}{2024}\natexlab{}.
\newblock \showarticletitle{GEO: Generative Engine Optimization}. In \bibinfo{booktitle}{\emph{Proceedings of the 30th ACM SIGKDD Conference on Knowledge Discovery and Data Mining}} (Barcelona, Spain) \emph{(\bibinfo{series}{KDD '24})}. \bibinfo{publisher}{Association for Computing Machinery}, \bibinfo{address}{New York, NY, USA}, \bibinfo{pages}{5–16}.
\newblock
\showISBNx{9798400704901}
\urldef\tempurl%
\url{https://doi.org/10.1145/3637528.3671900}
\showDOI{\tempurl}


\bibitem[Alaofi et~al\mbox{.}(2023)]%
        {llm-query-variations}
\bibfield{author}{\bibinfo{person}{Marwah Alaofi}, \bibinfo{person}{Luke Gallagher}, \bibinfo{person}{Mark Sanderson}, \bibinfo{person}{Falk Scholer}, {and} \bibinfo{person}{Paul Thomas}.} \bibinfo{year}{2023}\natexlab{}.
\newblock \showarticletitle{Can Generative LLMs Create Query Variants for Test Collections? An Exploratory Study}. In \bibinfo{booktitle}{\emph{Proceedings of the 46th International ACM SIGIR Conference on Research and Development in Information Retrieval}} (Taipei, Taiwan) \emph{(\bibinfo{series}{SIGIR '23})}. \bibinfo{publisher}{Association for Computing Machinery}, \bibinfo{address}{New York, NY, USA}, \bibinfo{pages}{1869–1873}.
\newblock
\showISBNx{9781450394086}
\urldef\tempurl%
\url{https://doi.org/10.1145/3539618.3591960}
\showDOI{\tempurl}


\bibitem[Bailey et~al\mbox{.}(2016)]%
        {uqv100}
\bibfield{author}{\bibinfo{person}{Peter Bailey}, \bibinfo{person}{Alistair Moffat}, \bibinfo{person}{Falk Scholer}, {and} \bibinfo{person}{Paul Thomas}.} \bibinfo{year}{2016}\natexlab{}.
\newblock \showarticletitle{UQV100: A Test Collection with Query Variability} \emph{(\bibinfo{series}{SIGIR '16})}. \bibinfo{publisher}{Association for Computing Machinery}, \bibinfo{address}{New York, NY, USA}, \bibinfo{pages}{725–728}.
\newblock
\showISBNx{9781450340694}
\urldef\tempurl%
\url{https://doi.org/10.1145/2911451.2914671}
\showDOI{\tempurl}


\bibitem[Balasubramanian and Allan(2010)]%
        {Balasubramanian+Allan:10a}
\bibfield{author}{\bibinfo{person}{Niranjan Balasubramanian} {and} \bibinfo{person}{James Allan}.} \bibinfo{year}{2010}\natexlab{}.
\newblock \showarticletitle{Learning to select rankers}. In \bibinfo{booktitle}{\emph{Proceeding of SIGIR}}. \bibinfo{pages}{855--856}.
\newblock


\bibitem[Balog and Zhai(2024)]%
        {Balog+Zhai:24a}
\bibfield{author}{\bibinfo{person}{Krisztian Balog} {and} \bibinfo{person}{ChengXiang Zhai}.} \bibinfo{year}{2024}\natexlab{}.
\newblock \showarticletitle{User Simulation for Evaluating Information Access Systems}.
\newblock \bibinfo{journal}{\emph{Found. Trends Inf. Retr.}} \bibinfo{volume}{18}, \bibinfo{number}{1-2} (\bibinfo{year}{2024}), \bibinfo{pages}{1--261}.
\newblock


\bibitem[Bar{-}Yossef and Gurevich(2008)]%
        {Yossef+Gurevich:08a}
\bibfield{author}{\bibinfo{person}{Ziv Bar{-}Yossef} {and} \bibinfo{person}{Maxim Gurevich}.} \bibinfo{year}{2008}\natexlab{}.
\newblock \showarticletitle{Random sampling from a search engine's index}.
\newblock \bibinfo{journal}{\emph{J. {ACM}}} \bibinfo{volume}{55}, \bibinfo{number}{5} (\bibinfo{year}{2008}), \bibinfo{pages}{24:1--24:74}.
\newblock


\bibitem[Bardas et~al\mbox{.}(2025)]%
        {Niv}
\bibfield{author}{\bibinfo{person}{Niv Bardas}, \bibinfo{person}{Tommy Mordo}, \bibinfo{person}{Oren Kurland}, \bibinfo{person}{Moshe Tennenholtz}, {and} \bibinfo{person}{Gal Zur}.} \bibinfo{year}{2025}\natexlab{}.
\newblock \bibinfo{title}{Prompt-Based Document Modifications In Ranking Competitions}.
\newblock
\newblock
\showeprint[arxiv]{2502.07315}~[cs.IR]
\urldef\tempurl%
\url{https://arxiv.org/abs/2502.07315}
\showURL{%
\tempurl}


\bibitem[Bender et~al\mbox{.}(2021)]%
        {llm-parrots}
\bibfield{author}{\bibinfo{person}{Emily~M. Bender}, \bibinfo{person}{Timnit Gebru}, \bibinfo{person}{Angelina McMillan-Major}, {and} \bibinfo{person}{Shmargaret Shmitchell}.} \bibinfo{year}{2021}\natexlab{}.
\newblock \showarticletitle{On the Dangers of Stochastic Parrots: Can Language Models Be Too Big?}. In \bibinfo{booktitle}{\emph{Proceedings of the 2021 ACM Conference on Fairness, Accountability, and Transparency}} (Virtual Event, Canada) \emph{(\bibinfo{series}{FAccT '21})}. \bibinfo{publisher}{Association for Computing Machinery}, \bibinfo{address}{New York, NY, USA}, \bibinfo{pages}{610–623}.
\newblock
\showISBNx{9781450383097}
\urldef\tempurl%
\url{https://doi.org/10.1145/3442188.3445922}
\showDOI{\tempurl}


\bibitem[Cabanac et~al\mbox{.}(2021)]%
        {tortured-phrases}
\bibfield{author}{\bibinfo{person}{Guillaume Cabanac}, \bibinfo{person}{Cyril Labbé}, {and} \bibinfo{person}{Alexander Magazinov}.} \bibinfo{year}{2021}\natexlab{}.
\newblock \bibinfo{title}{Tortured phrases: A dubious writing style emerging in science. Evidence of critical issues affecting established journals}.
\newblock
\newblock
\showeprint[arxiv]{2107.06751}~[cs.DL]
\urldef\tempurl%
\url{https://arxiv.org/abs/2107.06751}
\showURL{%
\tempurl}


\bibitem[Campos et~al\mbox{.}(2020)]%
        {YAKE}
\bibfield{author}{\bibinfo{person}{Ricardo Campos}, \bibinfo{person}{V\'{\i}tor Mangaravite}, \bibinfo{person}{Arian Pasquali}, \bibinfo{person}{Al\'{\i}pio Jorge}, \bibinfo{person}{C\'{e}lia Nunes}, {and} \bibinfo{person}{Adam Jatowt}.} \bibinfo{year}{2020}\natexlab{}.
\newblock \showarticletitle{YAKE! Keyword extraction from single documents using multiple local features}.
\newblock \bibinfo{journal}{\emph{Inf. Sci.}} \bibinfo{volume}{509}, \bibinfo{number}{C} (\bibinfo{date}{Jan.} \bibinfo{year}{2020}), \bibinfo{pages}{257–289}.
\newblock
\showISSN{0020-0255}
\urldef\tempurl%
\url{https://doi.org/10.1016/j.ins.2019.09.013}
\showDOI{\tempurl}


\bibitem[Chen et~al\mbox{.}(2024)]%
        {chen2024bge}
\bibfield{author}{\bibinfo{person}{Jianlv Chen}, \bibinfo{person}{Shitao Xiao}, \bibinfo{person}{Peitian Zhang}, \bibinfo{person}{Kun Luo}, \bibinfo{person}{Defu Lian}, {and} \bibinfo{person}{Zheng Liu}.} \bibinfo{year}{2024}\natexlab{}.
\newblock \bibinfo{title}{BGE M3-Embedding: Multi-Lingual, Multi-Functionality, Multi-Granularity Text Embeddings Through Self-Knowledge Distillation}.
\newblock
\newblock
\showeprint[arxiv]{2402.03216}~[cs.CL]


\bibitem[Chen et~al\mbox{.}(2023a)]%
        {Chen+al:23a}
\bibfield{author}{\bibinfo{person}{Xuanang Chen}, \bibinfo{person}{Ben He}, \bibinfo{person}{Le Sun}, {and} \bibinfo{person}{Yingfei Sun}.} \bibinfo{year}{2023}\natexlab{a}.
\newblock \showarticletitle{Defense of Adversarial Ranking Attack in Text Retrieval: Benchmark and Baseline via Detection}.
\newblock \bibinfo{journal}{\emph{CoRR}}  \bibinfo{volume}{abs/2307.16816} (\bibinfo{year}{2023}).
\newblock


\bibitem[Chen et~al\mbox{.}(2023b)]%
        {Chen+al:23b}
\bibfield{author}{\bibinfo{person}{Xuanang Chen}, \bibinfo{person}{Ben He}, \bibinfo{person}{Zheng Ye}, \bibinfo{person}{Le Sun}, {and} \bibinfo{person}{Yingfei Sun}.} \bibinfo{year}{2023}\natexlab{b}.
\newblock \showarticletitle{Towards Imperceptible Document Manipulations against Neural Ranking Models}. In \bibinfo{booktitle}{\emph{Findings of the Association for Computational Linguistics: {ACL}}}. \bibinfo{pages}{6648--6664}.
\newblock


\bibitem[Chen et~al\mbox{.}(2021)]%
        {Chen+al:21a}
\bibfield{author}{\bibinfo{person}{Zhumin Chen}, \bibinfo{person}{Xueqi Cheng}, \bibinfo{person}{Shoubin Dong}, \bibinfo{person}{Zhicheng Dou}, \bibinfo{person}{Jiafeng Guo}, \bibinfo{person}{Xuanjing Huang}, \bibinfo{person}{Yanyan Lan}, \bibinfo{person}{Chenliang Li}, \bibinfo{person}{Ru Li}, \bibinfo{person}{Tie{-}Yan Liu}, \bibinfo{person}{Yiqun Liu}, \bibinfo{person}{Jun Ma}, \bibinfo{person}{Bing Qin}, \bibinfo{person}{Mingwen Wang}, \bibinfo{person}{Ji{-}Rong Wen}, \bibinfo{person}{Jun Xu}, \bibinfo{person}{Min Zhang}, \bibinfo{person}{Peng Zhang}, {and} \bibinfo{person}{Qi Zhang}.} \bibinfo{year}{2021}\natexlab{}.
\newblock \showarticletitle{Information retrieval: a view from the Chinese {IR} community}.
\newblock \bibinfo{journal}{\emph{Frontiers Comput. Sci.}} \bibinfo{volume}{15}, \bibinfo{number}{1} (\bibinfo{year}{2021}), \bibinfo{pages}{151601}.
\newblock


\bibitem[Cheng and Fang(2020)]%
        {Cheng+Fang:20a}
\bibfield{author}{\bibinfo{person}{Zitong Cheng} {and} \bibinfo{person}{Hui Fang}.} \bibinfo{year}{2020}\natexlab{}.
\newblock \showarticletitle{Utilizing Axiomatic Perturbations to Guide Neural Ranking Models}. In \bibinfo{booktitle}{\emph{Proceedings of ICTIR}}. \bibinfo{pages}{153--156}.
\newblock


\bibitem[Cleverdon(1997)]%
        {cranfield}
\bibfield{author}{\bibinfo{person}{Cyril Cleverdon}.} \bibinfo{year}{1997}\natexlab{}.
\newblock \bibinfo{booktitle}{\emph{The Cranfield tests on index language devices}}.
\newblock \bibinfo{publisher}{Morgan Kaufmann Publishers Inc.}, \bibinfo{address}{San Francisco, CA, USA}, \bibinfo{pages}{47–59}.
\newblock
\showISBNx{1558604545}


\bibitem[Culpepper et~al\mbox{.}(2022)]%
        {shane:22a}
\bibfield{author}{\bibinfo{person}{J.~Shane Culpepper}, \bibinfo{person}{Guglielmo Faggioli}, \bibinfo{person}{Nicola Ferro}, {and} \bibinfo{person}{Oren Kurland}.} \bibinfo{year}{2022}\natexlab{}.
\newblock \showarticletitle{Topic Difficulty: Collection and Query Formulation Effects}.
\newblock \bibinfo{journal}{\emph{{ACM} Trans. Inf. Syst.}} \bibinfo{volume}{40}, \bibinfo{number}{1} (\bibinfo{year}{2022}), \bibinfo{pages}{19:1--19:36}.
\newblock


\bibitem[Cuskley et~al\mbox{.}(2024)]%
        {how-llms-understand-language}
\bibfield{author}{\bibinfo{person}{Christine Cuskley}, \bibinfo{person}{Rebecca Woods}, {and} \bibinfo{person}{Molly Flaherty}.} \bibinfo{year}{2024}\natexlab{}.
\newblock \showarticletitle{The Limitations of Large Language Models for Understanding Human Language and Cognition}.
\newblock \bibinfo{journal}{\emph{Open Mind}}  \bibinfo{volume}{8} (\bibinfo{date}{08} \bibinfo{year}{2024}), \bibinfo{pages}{1058--1083}.
\newblock
\showISSN{2470-2986}
\urldef\tempurl%
\url{https://doi.org/10.1162/opmi_a_00160}
\showDOI{\tempurl}
\showeprint{https://direct.mit.edu/opmi/article-pdf/doi/10.1162/opmi\_a\_00160/2468254/opmi\_a\_00160.pdf}


\bibitem[Dai et~al\mbox{.}(2024a)]%
        {dai2024cocktail}
\bibfield{author}{\bibinfo{person}{Sunhao Dai}, \bibinfo{person}{Weihao Liu}, \bibinfo{person}{Yuqi Zhou}, \bibinfo{person}{Liang Pang}, \bibinfo{person}{Rongju Ruan}, \bibinfo{person}{Gang Wang}, \bibinfo{person}{Zhenhua Dong}, \bibinfo{person}{Jun Xu}, {and} \bibinfo{person}{Ji-Rong Wen}.} \bibinfo{year}{2024}\natexlab{a}.
\newblock \showarticletitle{Cocktail: A Comprehensive Information Retrieval Benchmark with LLM-Generated Documents Integration}.
\newblock \bibinfo{journal}{\emph{Findings of the Association for Computational Linguistics: ACL 2024}} (\bibinfo{year}{2024}).
\newblock


\bibitem[Dai et~al\mbox{.}(2024b)]%
        {neural-llm-bias}
\bibfield{author}{\bibinfo{person}{Sunhao Dai}, \bibinfo{person}{Yuqi Zhou}, \bibinfo{person}{Liang Pang}, \bibinfo{person}{Weihao Liu}, \bibinfo{person}{Xiaolin Hu}, \bibinfo{person}{Yong Liu}, \bibinfo{person}{Xiao Zhang}, \bibinfo{person}{Gang Wang}, {and} \bibinfo{person}{Jun Xu}.} \bibinfo{year}{2024}\natexlab{b}.
\newblock \showarticletitle{Neural Retrievers are Biased Towards LLM-Generated Content}. In \bibinfo{booktitle}{\emph{Proceedings of the 30th ACM SIGKDD Conference on Knowledge Discovery and Data Mining}} (Barcelona, Spain) \emph{(\bibinfo{series}{KDD '24})}. \bibinfo{publisher}{Association for Computing Machinery}, \bibinfo{address}{New York, NY, USA}, \bibinfo{pages}{526–537}.
\newblock
\showISBNx{9798400704901}
\urldef\tempurl%
\url{https://doi.org/10.1145/3637528.3671882}
\showDOI{\tempurl}


\bibitem[Devlin et~al\mbox{.}(2018)]%
        {bert-transformer}
\bibfield{author}{\bibinfo{person}{Jacob Devlin}, \bibinfo{person}{Ming{-}Wei Chang}, \bibinfo{person}{Kenton Lee}, {and} \bibinfo{person}{Kristina Toutanova}.} \bibinfo{year}{2018}\natexlab{}.
\newblock \showarticletitle{{BERT:} Pre-training of Deep Bidirectional Transformers for Language Understanding}.
\newblock \bibinfo{journal}{\emph{CoRR}}  \bibinfo{volume}{abs/1810.04805} (\bibinfo{year}{2018}).
\newblock
\showeprint[arXiv]{1810.04805}
\urldef\tempurl%
\url{http://arxiv.org/abs/1810.04805}
\showURL{%
\tempurl}


\bibitem[Dubey et~al\mbox{.}(2024)]%
        {dubey_llama_2024}
\bibfield{author}{\bibinfo{person}{Abhimanyu Dubey}, \bibinfo{person}{Abhinav Jauhri}, \bibinfo{person}{Abhinav Pandey}, \bibinfo{person}{Abhishek Kadian}, \bibinfo{person}{Al-Dahle}, {et~al\mbox{.}}} \bibinfo{year}{2024}\natexlab{}.
\newblock \bibinfo{title}{The {Llama} 3 {Herd} of {Models}}.
\newblock
\newblock
\urldef\tempurl%
\url{http://arxiv.org/abs/2407.21783}
\showURL{%
\tempurl}
\newblock
\shownote{arXiv:2407.21783 [cs]}.


\bibitem[Efron et~al\mbox{.}(2012)]%
        {Efron+al:12a}
\bibfield{author}{\bibinfo{person}{Miles Efron}, \bibinfo{person}{Peter Organisciak}, {and} \bibinfo{person}{Katrina Fenlon}.} \bibinfo{year}{2012}\natexlab{}.
\newblock \showarticletitle{Improving retrieval of short texts through document expansion}. In \bibinfo{booktitle}{\emph{Proceedings of SIGIR}}. \bibinfo{pages}{911--920}.
\newblock


\bibitem[Fang and Zhai(2005)]%
        {Fang+Zhai:05a}
\bibfield{author}{\bibinfo{person}{Hui Fang} {and} \bibinfo{person}{ChengXiang Zhai}.} \bibinfo{year}{2005}\natexlab{}.
\newblock \showarticletitle{An exploration of axiomatic approaches to information retrieval}. In \bibinfo{booktitle}{\emph{Proceedings of SIGIR}}. \bibinfo{pages}{480--487}.
\newblock


\bibitem[Galusc{\'{a}}kov{\'{a}} et~al\mbox{.}(2024)]%
        {crossling}
\bibfield{author}{\bibinfo{person}{Petra Galusc{\'{a}}kov{\'{a}}}, \bibinfo{person}{Douglas~W. Oard}, {and} \bibinfo{person}{Suraj Nair}.} \bibinfo{year}{2024}\natexlab{}.
\newblock \showarticletitle{Cross-language Retrieval}.
\newblock In \bibinfo{booktitle}{\emph{Information Retrieval: Advanced Topics and Techniques}}. \bibinfo{series}{{ACM} Books}, Vol.~\bibinfo{volume}{60}. \bibinfo{pages}{321--357}.
\newblock


\bibitem[{Gemma Team} et~al\mbox{.}({[n.\,d.]})]%
        {gemma_team_gemma_nodate}
\bibfield{author}{\bibinfo{person}{{Gemma Team}}, \bibinfo{person}{{Thomas Mesnard}}, \bibinfo{person}{{Cassidy Hardin}}, \bibinfo{person}{{Robert Dadashi}}, \bibinfo{person}{{Surya Bhupatiraju}}, \bibinfo{person}{{Laurent Sifre}}, \bibinfo{person}{{Morgane Rivière}}, \bibinfo{person}{{Mihir Sanjay Kale}}, {et~al\mbox{.}}} \bibinfo{year}{[n.\,d.]}\natexlab{}.
\newblock \bibinfo{title}{Gemma}.
\newblock
\newblock
\urldef\tempurl%
\url{https://doi.org/10.34740/KAGGLE/M/3301}
\showDOI{\tempurl}


\bibitem[Goren et~al\mbox{.}(2020)]%
        {Greg-bot}
\bibfield{author}{\bibinfo{person}{Gregory Goren}, \bibinfo{person}{Oren Kurland}, \bibinfo{person}{Moshe Tennenholtz}, {and} \bibinfo{person}{Fiana Raiber}.} \bibinfo{year}{2020}\natexlab{}.
\newblock \showarticletitle{Ranking-Incentivized Quality Preserving Content Modification}. In \bibinfo{booktitle}{\emph{Proceedings of the 43rd International ACM SIGIR Conference on Research and Development in Information Retrieval}} (Virtual Event, China) \emph{(\bibinfo{series}{SIGIR '20})}. \bibinfo{publisher}{Association for Computing Machinery}, \bibinfo{address}{New York, NY, USA}, \bibinfo{pages}{259–268}.
\newblock
\showISBNx{9781450380164}
\urldef\tempurl%
\url{https://doi.org/10.1145/3397271.3401058}
\showDOI{\tempurl}


\bibitem[Goren et~al\mbox{.}(2021)]%
        {Greg-Herding}
\bibfield{author}{\bibinfo{person}{Gregory Goren}, \bibinfo{person}{Oren Kurland}, \bibinfo{person}{Moshe Tennenholtz}, {and} \bibinfo{person}{Fiana Raiber}.} \bibinfo{year}{2021}\natexlab{}.
\newblock \showarticletitle{Driving the Herd: Search Engines as Content Influencers}. In \bibinfo{booktitle}{\emph{Proceedings of the 30th ACM International Conference on Information \& Knowledge Management}} (Virtual Event, Queensland, Australia) \emph{(\bibinfo{series}{CIKM '21})}. \bibinfo{publisher}{Association for Computing Machinery}, \bibinfo{address}{New York, NY, USA}, \bibinfo{pages}{586–595}.
\newblock
\showISBNx{9781450384469}
\urldef\tempurl%
\url{https://doi.org/10.1145/3459637.3482334}
\showDOI{\tempurl}


\bibitem[Gy{\"o}ngyi and Garcia-Molina(2005)]%
        {Gyongyi+Molina:05a}
\bibfield{author}{\bibinfo{person}{Zolt{\'a}n Gy{\"o}ngyi} {and} \bibinfo{person}{Hector Garcia-Molina}.} \bibinfo{year}{2005}\natexlab{}.
\newblock \showarticletitle{Web Spam Taxonomy}. In \bibinfo{booktitle}{\emph{Proceedings of AIRWeb 2005, First International Workshop on Adversarial Information Retrieval on the Web}}. \bibinfo{pages}{39--47}.
\newblock


\bibitem[Haider et~al\mbox{.}(2024)]%
        {gpt-papers}
\bibfield{author}{\bibinfo{person}{Jutta Haider}, \bibinfo{person}{Kristofer Söderström}, \bibinfo{person}{Björn Ekström}, {and} \bibinfo{person}{Malte Rödl}.} \bibinfo{year}{2024}\natexlab{}.
\newblock \showarticletitle{GPT-fabricated scientific papers on Google Scholar: Key features, spread, and implications for preempting evidence manipulation}.
\newblock   \bibinfo{volume}{5} (\bibinfo{date}{09} \bibinfo{year}{2024}).
\newblock
\urldef\tempurl%
\url{https://doi.org/10.37016/mr-2020-156}
\showDOI{\tempurl}


\bibitem[Harman and Voorhees(2006)]%
        {Harman+Voorhees:06a}
\bibfield{author}{\bibinfo{person}{Donna Harman} {and} \bibinfo{person}{Ellen~M. Voorhees}.} \bibinfo{year}{2006}\natexlab{}.
\newblock \showarticletitle{{TREC:} An overview}.
\newblock \bibinfo{journal}{\emph{Annu. Rev. Inf. Sci. Technol.}} \bibinfo{volume}{40}, \bibinfo{number}{1} (\bibinfo{year}{2006}), \bibinfo{pages}{113--155}.
\newblock


\bibitem[Honovich et~al\mbox{.}(2022)]%
        {NLI}
\bibfield{author}{\bibinfo{person}{Or Honovich}, \bibinfo{person}{Roee Aharoni}, \bibinfo{person}{Jonathan Herzig}, \bibinfo{person}{Hagai Taitelbaum}, \bibinfo{person}{Doron Kukliansy}, \bibinfo{person}{Vered Cohen}, \bibinfo{person}{Thomas Scialom}, \bibinfo{person}{Idan Szpektor}, \bibinfo{person}{Avinatan Hassidim}, {and} \bibinfo{person}{Yossi Matias}.} \bibinfo{year}{2022}\natexlab{}.
\newblock \showarticletitle{{TRUE}: Re-evaluating Factual Consistency Evaluation}. In \bibinfo{booktitle}{\emph{Proceedings of the 2022 Conference of the North American Chapter of the Association for Computational Linguistics: Human Language Technologies}}. \bibinfo{publisher}{Association for Computational Linguistics}, \bibinfo{address}{Seattle, United States}, \bibinfo{pages}{3905--3920}.
\newblock
\urldef\tempurl%
\url{https://doi.org/10.18653/v1/2022.naacl-main.287}
\showDOI{\tempurl}


\bibitem[Imasaka and Joho(2024)]%
        {query-generation-personality}
\bibfield{author}{\bibinfo{person}{Yuta Imasaka} {and} \bibinfo{person}{Hideo Joho}.} \bibinfo{year}{2024}\natexlab{}.
\newblock \showarticletitle{Effect of LLM's Personality Traits on Query Generation}. In \bibinfo{booktitle}{\emph{Proceedings of the 2024 Annual International ACM SIGIR Conference on Research and Development in Information Retrieval in the Asia Pacific Region}} (Tokyo, Japan) \emph{(\bibinfo{series}{SIGIR-AP 2024})}. \bibinfo{publisher}{Association for Computing Machinery}, \bibinfo{address}{New York, NY, USA}, \bibinfo{pages}{249–258}.
\newblock
\showISBNx{9798400707247}
\urldef\tempurl%
\url{https://doi.org/10.1145/3673791.3698433}
\showDOI{\tempurl}


\bibitem[Izacard et~al\mbox{.}(2022)]%
        {contriever}
\bibfield{author}{\bibinfo{person}{Gautier Izacard}, \bibinfo{person}{Mathilde Caron}, \bibinfo{person}{Lucas Hosseini}, \bibinfo{person}{Sebastian Riedel}, \bibinfo{person}{Piotr Bojanowski}, \bibinfo{person}{Armand Joulin}, {and} \bibinfo{person}{Edouard Grave}.} \bibinfo{year}{2022}\natexlab{}.
\newblock \bibinfo{title}{Unsupervised Dense Information Retrieval with Contrastive Learning}.
\newblock
\newblock
\showeprint[arxiv]{2112.09118}~[cs.IR]
\urldef\tempurl%
\url{https://arxiv.org/abs/2112.09118}
\showURL{%
\tempurl}


\bibitem[Jagerman et~al\mbox{.}(2023)]%
        {Jagerman+al:23a}
\bibfield{author}{\bibinfo{person}{Rolf Jagerman}, \bibinfo{person}{Honglei Zhuang}, \bibinfo{person}{Zhen Qin}, \bibinfo{person}{Xuanhui Wang}, {and} \bibinfo{person}{Michael Bendersky}.} \bibinfo{year}{2023}\natexlab{}.
\newblock \bibinfo{title}{Query Expansion by Prompting Large Language Models}.
\newblock
\newblock
\showeprint[arxiv]{2305.03653}~[cs.IR]


\bibitem[Jansen et~al\mbox{.}(2007)]%
        {Jansen+al:07a}
\bibfield{author}{\bibinfo{person}{Bernard~J. Jansen}, \bibinfo{person}{Danielle~L. Booth}, {and} \bibinfo{person}{Amanda Spink}.} \bibinfo{year}{2007}\natexlab{}.
\newblock \showarticletitle{Determining the user intent of web search engine queries}. In \bibinfo{booktitle}{\emph{Proceedings of WWW}}. \bibinfo{pages}{1149--1150}.
\newblock


\bibitem[Kurland and Culpepper(2018)]%
        {Kurland+Culpepper:18a}
\bibfield{author}{\bibinfo{person}{Oren Kurland} {and} \bibinfo{person}{J.~Shane Culpepper}.} \bibinfo{year}{2018}\natexlab{}.
\newblock \showarticletitle{Fusion in Information Retrieval: {SIGIR} 2018 Half-Day Tutorial}. In \bibinfo{booktitle}{\emph{Proceedings of SIGIR}}. \bibinfo{pages}{1383--1386}.
\newblock


\bibitem[Kurland and Lee(2004)]%
        {Kurland+Lee:04a}
\bibfield{author}{\bibinfo{person}{Oren Kurland} {and} \bibinfo{person}{Lillian Lee}.} \bibinfo{year}{2004}\natexlab{}.
\newblock \showarticletitle{Corpus structure, language models, and ad hoc information retrieval}. In \bibinfo{booktitle}{\emph{Proceedings of SIGIR}}. \bibinfo{pages}{194--201}.
\newblock


\bibitem[Kurland and Tennenholtz(2022)]%
        {kurland_competitive_2022}
\bibfield{author}{\bibinfo{person}{Oren Kurland} {and} \bibinfo{person}{Moshe Tennenholtz}.} \bibinfo{year}{2022}\natexlab{}.
\newblock \showarticletitle{Competitive {Search}}. In \bibinfo{booktitle}{\emph{Proceedings of SIGIR}}. \bibinfo{pages}{2838--2849}.
\newblock


\bibitem[Lafferty and Zhai(2001)]%
        {Lafferty+Zhai:01a}
\bibfield{author}{\bibinfo{person}{John~D. Lafferty} {and} \bibinfo{person}{Chengxiang Zhai}.} \bibinfo{year}{2001}\natexlab{}.
\newblock \showarticletitle{Document language models, query models, and risk minimization for information retrieval}. In \bibinfo{booktitle}{\emph{Proceedings of SIGIR}}. \bibinfo{pages}{111--119}.
\newblock


\bibitem[Lavrenko(2008)]%
        {lavrenko-rm-book}
\bibfield{author}{\bibinfo{person}{Victor Lavrenko}.} \bibinfo{year}{2008}\natexlab{}.
\newblock \bibinfo{booktitle}{\emph{A Generative Theory of Relevance} (\bibinfo{edition}{1st} ed.)}.
\newblock \bibinfo{publisher}{Springer Publishing Company, Incorporated}.
\newblock
\showISBNx{3540893636}


\bibitem[Lavrenko et~al\mbox{.}(2002)]%
        {Lavrenko+al:02a}
\bibfield{author}{\bibinfo{person}{Victor Lavrenko}, \bibinfo{person}{Martin Choquette}, {and} \bibinfo{person}{W.~Bruce Croft}.} \bibinfo{year}{2002}\natexlab{}.
\newblock \showarticletitle{Cross-lingual relevance models}. In \bibinfo{booktitle}{\emph{Proceedings of SIGIR}}. \bibinfo{pages}{175--182}.
\newblock


\bibitem[Lavrenko and Croft(2001)]%
        {Lavrenko+Croft:01a}
\bibfield{author}{\bibinfo{person}{Victor Lavrenko} {and} \bibinfo{person}{W.~Bruce Croft}.} \bibinfo{year}{2001}\natexlab{}.
\newblock \showarticletitle{Relevance-Based Language Models}. In \bibinfo{booktitle}{\emph{Proceedings of SIGIR}}. \bibinfo{pages}{120--127}.
\newblock


\bibitem[Liang et~al\mbox{.}(2022)]%
        {liang2022holistic}
\bibfield{author}{\bibinfo{person}{Percy Liang}, \bibinfo{person}{Rishi Bommasani}, \bibinfo{person}{Tony Lee}, \bibinfo{person}{Dimitris Tsipras}, \bibinfo{person}{Dilara Soylu}, \bibinfo{person}{Michihiro Yasunaga}, \bibinfo{person}{Yian Zhang}, \bibinfo{person}{Deepak Narayanan}, \bibinfo{person}{Yuhuai Wu}, \bibinfo{person}{Ananya Kumar}, \bibinfo{person}{Benjamin Newman}, \bibinfo{person}{Binhang Yuan}, \bibinfo{person}{Bobby Yan}, \bibinfo{person}{Ce Zhang}, \bibinfo{person}{Christian Cosgrove}, \bibinfo{person}{Christopher~D. Manning}, \bibinfo{person}{Christopher R'e}, \bibinfo{person}{Diana Acosta-Navas}, \bibinfo{person}{Drew~A. Hudson}, \bibinfo{person}{E. Zelikman}, \bibinfo{person}{Esin Durmus}, \bibinfo{person}{Faisal Ladhak}, \bibinfo{person}{Frieda Rong}, \bibinfo{person}{Hongyu Ren}, \bibinfo{person}{Huaxiu Yao}, \bibinfo{person}{Jue Wang}, \bibinfo{person}{Keshav Santhanam}, \bibinfo{person}{Laurel~J. Orr}, \bibinfo{person}{Lucia Zheng}, \bibinfo{person}{Mert Yuksekgonul},
  \bibinfo{person}{Mirac Suzgun}, \bibinfo{person}{Nathan~S. Kim}, \bibinfo{person}{Neel Guha}, \bibinfo{person}{Niladri~S. Chatterji}, \bibinfo{person}{O. Khattab}, \bibinfo{person}{Peter Henderson}, \bibinfo{person}{Qian Huang}, \bibinfo{person}{Ryan Chi}, \bibinfo{person}{Sang~Michael Xie}, \bibinfo{person}{Shibani Santurkar}, \bibinfo{person}{Surya Ganguli}, \bibinfo{person}{Tatsunori Hashimoto}, \bibinfo{person}{Thomas~F. Icard}, \bibinfo{person}{Tianyi Zhang}, \bibinfo{person}{Vishrav Chaudhary}, \bibinfo{person}{William Wang}, \bibinfo{person}{Xuechen Li}, \bibinfo{person}{Yifan Mai}, \bibinfo{person}{Yuhui Zhang}, {and} \bibinfo{person}{Yuta Koreeda}.} \bibinfo{year}{2022}\natexlab{}.
\newblock \showarticletitle{Holistic evaluation of language models}.
\newblock \bibinfo{journal}{\emph{arXiv}}  \bibinfo{volume}{abs/2211.09110} (\bibinfo{year}{2022}).
\newblock
\urldef\tempurl%
\url{https://arxiv.org/abs/2211.09110}
\showURL{%
\tempurl}


\bibitem[Liang et~al\mbox{.}(2024)]%
        {conference-llm-reviews}
\bibfield{author}{\bibinfo{person}{Weixin Liang}, \bibinfo{person}{Zachary Izzo}, \bibinfo{person}{Yaohui Zhang}, \bibinfo{person}{Haley Lepp}, \bibinfo{person}{Hancheng Cao}, \bibinfo{person}{Xuandong Zhao}, \bibinfo{person}{Lingjiao Chen}, \bibinfo{person}{Haotian Ye}, \bibinfo{person}{Sheng Liu}, \bibinfo{person}{Zhi Huang}, {et~al\mbox{.}}} \bibinfo{year}{2024}\natexlab{}.
\newblock \showarticletitle{Monitoring AI-Modified Content at Scale: A Case Study on the Impact of ChatGPT on AI Conference Peer Reviews}.
\newblock \bibinfo{journal}{\emph{arXiv preprint arXiv:2403.07183}} (\bibinfo{year}{2024}).
\newblock


\bibitem[Liu et~al\mbox{.}(2022)]%
        {Liu+al:22a}
\bibfield{author}{\bibinfo{person}{Jiawei Liu}, \bibinfo{person}{Yangyang Kang}, \bibinfo{person}{Di Tang}, \bibinfo{person}{Kaisong Song}, \bibinfo{person}{Changlong Sun}, \bibinfo{person}{Xiaofeng Wang}, \bibinfo{person}{Wei Lu}, {and} \bibinfo{person}{Xiaozhong Liu}.} \bibinfo{year}{2022}\natexlab{}.
\newblock \showarticletitle{Order-Disorder: Imitation Adversarial Attacks for Black-box Neural Ranking Models}. In \bibinfo{booktitle}{\emph{Proceedings of {SIGSAC}}}. \bibinfo{pages}{2025--2039}.
\newblock


\bibitem[Liu and Croft(2004)]%
        {Liu+Croft:04a}
\bibfield{author}{\bibinfo{person}{Xiaoyong Liu} {and} \bibinfo{person}{W.~Bruce Croft}.} \bibinfo{year}{2004}\natexlab{}.
\newblock \showarticletitle{Cluster-Based Retrieval Using Language Models}. In \bibinfo{booktitle}{\emph{Proceedings of SIGIR}}. \bibinfo{pages}{186--193}.
\newblock


\bibitem[Ma et~al\mbox{.}(2023)]%
        {query-rewrite-for-rag}
\bibfield{author}{\bibinfo{person}{Xinbei Ma}, \bibinfo{person}{Yeyun Gong}, \bibinfo{person}{Pengcheng He}, \bibinfo{person}{Hai Zhao}, {and} \bibinfo{person}{Nan Duan}.} \bibinfo{year}{2023}\natexlab{}.
\newblock \bibinfo{title}{Query Rewriting for Retrieval-Augmented Large Language Models}.
\newblock
\newblock
\showeprint[arxiv]{2305.14283}~[cs.CL]
\urldef\tempurl%
\url{https://arxiv.org/abs/2305.14283}
\showURL{%
\tempurl}


\bibitem[Mordo et~al\mbox{.}(2025)]%
        {Lemss}
\bibfield{author}{\bibinfo{person}{Tommy Mordo}, \bibinfo{person}{Tomer Kordonsky}, \bibinfo{person}{Haya Nachimovsky}, \bibinfo{person}{Moshe Tennenholtz}, {and} \bibinfo{person}{Oren Kurland}.} \bibinfo{year}{2025}\natexlab{}.
\newblock \bibinfo{title}{CSP: A Simulator For Multi-Agent Ranking Competitions}.
\newblock
\newblock
\showeprint[arxiv]{2502.11197}~[cs.IR]
\urldef\tempurl%
\url{https://arxiv.org/abs/2502.11197}
\showURL{%
\tempurl}


\bibitem[Nachimovsky et~al\mbox{.}(2024)]%
        {MultQueries}
\bibfield{author}{\bibinfo{person}{Haya Nachimovsky}, \bibinfo{person}{Moshe Tennenholtz}, \bibinfo{person}{Fiana Raiber}, {and} \bibinfo{person}{Oren Kurland}.} \bibinfo{year}{2024}\natexlab{}.
\newblock \showarticletitle{Ranking-Incentivized Document Manipulations for Multiple Queries}. In \bibinfo{booktitle}{\emph{Proceedings of the 2024 ACM SIGIR International Conference on Theory of Information Retrieval}} (Washington DC, USA) \emph{(\bibinfo{series}{ICTIR '24})}. \bibinfo{publisher}{Association for Computing Machinery}, \bibinfo{address}{New York, NY, USA}, \bibinfo{pages}{61–70}.
\newblock
\showISBNx{9798400706813}
\urldef\tempurl%
\url{https://doi.org/10.1145/3664190.3672516}
\showDOI{\tempurl}


\bibitem[Nguyen et~al\mbox{.}(2016)]%
        {msmarco}
\bibfield{author}{\bibinfo{person}{Tri Nguyen}, \bibinfo{person}{Mir Rosenberg}, \bibinfo{person}{Xia Song}, \bibinfo{person}{Jianfeng Gao}, \bibinfo{person}{Saurabh Tiwary}, \bibinfo{person}{Rangan Majumder}, {and} \bibinfo{person}{Li Deng}.} \bibinfo{year}{2016}\natexlab{}.
\newblock \showarticletitle{{MS} {MARCO:} {A} Human Generated MAchine Reading COmprehension Dataset}.
\newblock \bibinfo{journal}{\emph{CoRR}}  \bibinfo{volume}{abs/1611.09268} (\bibinfo{year}{2016}).
\newblock
\showeprint[arXiv]{1611.09268}
\urldef\tempurl%
\url{http://arxiv.org/abs/1611.09268}
\showURL{%
\tempurl}


\bibitem[Nogueira et~al\mbox{.}(2019a)]%
        {Nogueira+al:19a}
\bibfield{author}{\bibinfo{person}{Rodrigo Nogueira}, \bibinfo{person}{Wei Yang}, \bibinfo{person}{Jimmy Lin}, {and} \bibinfo{person}{Kyunghyun Cho}.} \bibinfo{year}{2019}\natexlab{a}.
\newblock \bibinfo{title}{Document Expansion by Query Prediction}.
\newblock
\newblock
\showeprint[arxiv]{1904.08375}~[cs.IR]


\bibitem[Nogueira and Cho(2019)]%
        {BERT}
\bibfield{author}{\bibinfo{person}{Rodrigo~Frassetto Nogueira} {and} \bibinfo{person}{Kyunghyun Cho}.} \bibinfo{year}{2019}\natexlab{}.
\newblock \showarticletitle{Passage Re-ranking with {BERT}}.
\newblock \bibinfo{journal}{\emph{CoRR}}  \bibinfo{volume}{abs/1901.04085} (\bibinfo{year}{2019}).
\newblock
\showeprint[arXiv]{1901.04085}
\urldef\tempurl%
\url{http://arxiv.org/abs/1901.04085}
\showURL{%
\tempurl}


\bibitem[Nogueira et~al\mbox{.}(2019b)]%
        {doc2query}
\bibfield{author}{\bibinfo{person}{Rodrigo~Frassetto Nogueira}, \bibinfo{person}{Wei Yang}, \bibinfo{person}{Jimmy Lin}, {and} \bibinfo{person}{Kyunghyun Cho}.} \bibinfo{year}{2019}\natexlab{b}.
\newblock \showarticletitle{Document Expansion by Query Prediction}.
\newblock \bibinfo{journal}{\emph{CoRR}}  \bibinfo{volume}{abs/1904.08375} (\bibinfo{year}{2019}).
\newblock
\showeprint[arXiv]{1904.08375}
\urldef\tempurl%
\url{http://arxiv.org/abs/1904.08375}
\showURL{%
\tempurl}


\bibitem[Pfrommer et~al\mbox{.}(2024)]%
        {ranking-manipulation-conversational}
\bibfield{author}{\bibinfo{person}{Samuel Pfrommer}, \bibinfo{person}{Yatong Bai}, \bibinfo{person}{Tanmay Gautam}, {and} \bibinfo{person}{Somayeh Sojoudi}.} \bibinfo{year}{2024}\natexlab{}.
\newblock \bibinfo{title}{Ranking Manipulation for Conversational Search Engines}.
\newblock
\newblock
\showeprint[arxiv]{2406.03589}~[cs.CL]
\urldef\tempurl%
\url{https://arxiv.org/abs/2406.03589}
\showURL{%
\tempurl}


\bibitem[Qin et~al\mbox{.}(2024)]%
        {qin2024pairwise}
\bibfield{author}{\bibinfo{person}{Zhen Qin}, \bibinfo{person}{Rolf Jagerman}, \bibinfo{person}{Kai Hui}, \bibinfo{person}{Honglei Zhuang}, \bibinfo{person}{Junru Wu}, \bibinfo{person}{Le Yan}, \bibinfo{person}{Jiaming Shen}, \bibinfo{person}{Tianqi Liu}, \bibinfo{person}{Jialu Liu}, \bibinfo{person}{Donald Metzler}, \bibinfo{person}{Xuanhui Wang}, {and} \bibinfo{person}{Michael Bendersky}.} \bibinfo{year}{2024}\natexlab{}.
\newblock \showarticletitle{Large Language Models are Effective Text Rankers with Pairwise Ranking Prompting}.
\newblock \bibinfo{journal}{\emph{arXiv}} (\bibinfo{year}{2024}).
\newblock
\urldef\tempurl%
\url{https://doi.org/10.48550/arXiv.2306.17563}
\showDOI{\tempurl}


\bibitem[Radinsky et~al\mbox{.}(2013)]%
        {temporal-web-dynamics}
\bibfield{author}{\bibinfo{person}{Kira Radinsky}, \bibinfo{person}{Fernando Diaz}, \bibinfo{person}{Susan Dumais}, \bibinfo{person}{Milad Shokouhi}, \bibinfo{person}{Anlei Dong}, {and} \bibinfo{person}{Yi Chang}.} \bibinfo{year}{2013}\natexlab{}.
\newblock \showarticletitle{Temporal web dynamics and its application to information retrieval}. In \bibinfo{booktitle}{\emph{Proceedings of the Sixth ACM International Conference on Web Search and Data Mining}} (Rome, Italy) \emph{(\bibinfo{series}{WSDM '13})}. \bibinfo{publisher}{Association for Computing Machinery}, \bibinfo{address}{New York, NY, USA}, \bibinfo{pages}{781–782}.
\newblock
\showISBNx{9781450318693}
\urldef\tempurl%
\url{https://doi.org/10.1145/2433396.2433500}
\showDOI{\tempurl}


\bibitem[Raifer et~al\mbox{.}(2017)]%
        {Nimrod}
\bibfield{author}{\bibinfo{person}{Nimrod Raifer}, \bibinfo{person}{Fiana Raiber}, \bibinfo{person}{Moshe Tennenholtz}, {and} \bibinfo{person}{Oren Kurland}.} \bibinfo{year}{2017}\natexlab{}.
\newblock \showarticletitle{Information Retrieval Meets Game Theory: The Ranking Competition Between Documents' Authors}. In \bibinfo{booktitle}{\emph{Proceedings of the 40th International ACM SIGIR Conference on Research and Development in Information Retrieval}} (Shinjuku, Tokyo, Japan) \emph{(\bibinfo{series}{SIGIR '17})}. \bibinfo{publisher}{Association for Computing Machinery}, \bibinfo{address}{New York, NY, USA}, \bibinfo{pages}{465–474}.
\newblock
\showISBNx{9781450350228}
\urldef\tempurl%
\url{https://doi.org/10.1145/3077136.3080785}
\showDOI{\tempurl}


\bibitem[Rashidi et~al\mbox{.}(2024)]%
        {Rashidi+al:24a}
\bibfield{author}{\bibinfo{person}{Lida Rashidi}, \bibinfo{person}{Justin Zobel}, {and} \bibinfo{person}{Alistair Moffat}.} \bibinfo{year}{2024}\natexlab{}.
\newblock \showarticletitle{Query Variability and Experimental Consistency: {A} Concerning Case Study}. In \bibinfo{booktitle}{\emph{Proceedings of ICTIR}}. \bibinfo{pages}{35--41}.
\newblock


\bibitem[Robertson(1977)]%
        {Robertson:77a}
\bibfield{author}{\bibinfo{person}{Stephen~E. Robertson}.} \bibinfo{year}{1977}\natexlab{}.
\newblock \showarticletitle{The Probability Ranking Principle in {IR}}.
\newblock \bibinfo{journal}{\emph{Journal of Documentation}} (\bibinfo{year}{1977}), \bibinfo{pages}{294--304}.
\newblock
\newblock
\shownote{Reprinted in K. Sparck Jones and P. Willett (eds), {\it Readings in Information Retrieval}, pp. 281--286, 1997}.


\bibitem[Robertson et~al\mbox{.}(1994)]%
        {Robertson:93a}
\bibfield{author}{\bibinfo{person}{Stephen~E. Robertson}, \bibinfo{person}{Steve Walker}, \bibinfo{person}{Susan Jones}, \bibinfo{person}{Micheline Hancock-Beaulieu}, {and} \bibinfo{person}{Mike Gatford}.} \bibinfo{year}{1994}\natexlab{}.
\newblock \showarticletitle{Okapi at TREC-3}. In \bibinfo{booktitle}{\emph{Proceedings of {TREC-3}}}.
\newblock


\bibitem[Shah and White(2024)]%
        {Shah+White:24a}
\bibfield{author}{\bibinfo{person}{Chirag Shah} {and} \bibinfo{person}{Ryen~W. White}.} \bibinfo{year}{2024}\natexlab{}.
\newblock \showarticletitle{Agents Are Not Enough}.
\newblock \bibinfo{journal}{\emph{CoRR}}  \bibinfo{volume}{abs/2412.16241} (\bibinfo{year}{2024}).
\newblock


\bibitem[Singhal and Kaszkiel(2001)]%
        {Singhal+Kaszkiel:01a}
\bibfield{author}{\bibinfo{person}{Amit Singhal} {and} \bibinfo{person}{Marcin Kaszkiel}.} \bibinfo{year}{2001}\natexlab{}.
\newblock \showarticletitle{A case study in web search using {TREC} algorithms}. In \bibinfo{booktitle}{\emph{Proceedings of {WWW}}}. \bibinfo{pages}{708--716}.
\newblock


\bibitem[Song et~al\mbox{.}(2022)]%
        {Song+al:22a}
\bibfield{author}{\bibinfo{person}{Junshuai Song}, \bibinfo{person}{Jiangshan Zhang}, \bibinfo{person}{Jifeng Zhu}, \bibinfo{person}{Mengyun Tang}, {and} \bibinfo{person}{Yong Yang}.} \bibinfo{year}{2022}\natexlab{}.
\newblock \showarticletitle{TRAttack: Text Rewriting Attack Against Text Retrieval}. In \bibinfo{booktitle}{\emph{Proceedings of RepL4NLP@ACL}}. \bibinfo{pages}{191--203}.
\newblock


\bibitem[Sun et~al\mbox{.}(2024)]%
        {ai-generated-social-media}
\bibfield{author}{\bibinfo{person}{Zhen Sun}, \bibinfo{person}{Zongmin Zhang}, \bibinfo{person}{Xinyue Shen}, \bibinfo{person}{Ziyi Zhang}, \bibinfo{person}{Yule Liu}, \bibinfo{person}{Michael Backes}, \bibinfo{person}{Yang Zhang}, {and} \bibinfo{person}{Xinlei He}.} \bibinfo{year}{2024}\natexlab{}.
\newblock \bibinfo{title}{Are We in the AI-Generated Text World Already? Quantifying and Monitoring AIGT on Social Media}.
\newblock
\newblock
\showeprint[arxiv]{2412.18148}~[cs.AI]
\urldef\tempurl%
\url{https://arxiv.org/abs/2412.18148}
\showURL{%
\tempurl}


\bibitem[Vasilisky et~al\mbox{.}(2023)]%
        {Ziv-Ranker}
\bibfield{author}{\bibinfo{person}{Ziv Vasilisky}, \bibinfo{person}{Oren Kurland}, \bibinfo{person}{Moshe Tennenholtz}, {and} \bibinfo{person}{Fiana Raiber}.} \bibinfo{year}{2023}\natexlab{}.
\newblock \showarticletitle{Content-Based Relevance Estimation in Retrieval Settings with Ranking-Incentivized Document Manipulations}. In \bibinfo{booktitle}{\emph{Proceedings of the 2023 ACM SIGIR International Conference on Theory of Information Retrieval}} (Taipei, Taiwan) \emph{(\bibinfo{series}{ICTIR '23})}. \bibinfo{publisher}{Association for Computing Machinery}, \bibinfo{address}{New York, NY, USA}, \bibinfo{pages}{205–214}.
\newblock
\showISBNx{9798400700736}
\urldef\tempurl%
\url{https://doi.org/10.1145/3578337.3605124}
\showDOI{\tempurl}


\bibitem[Wang et~al\mbox{.}(2022b)]%
        {e5}
\bibfield{author}{\bibinfo{person}{Liang Wang}, \bibinfo{person}{Nan Yang}, \bibinfo{person}{Xiaolong Huang}, \bibinfo{person}{Binxing Jiao}, \bibinfo{person}{Linjun Yang}, \bibinfo{person}{Daxin Jiang}, \bibinfo{person}{Rangan Majumder}, {and} \bibinfo{person}{Furu Wei}.} \bibinfo{year}{2022}\natexlab{b}.
\newblock \showarticletitle{Text Embeddings by Weakly-Supervised Contrastive Pre-training}.
\newblock \bibinfo{journal}{\emph{arXiv}} (\bibinfo{year}{2022}).
\newblock
\urldef\tempurl%
\url{https://doi.org/10.48550/arXiv.2212.03533}
\showDOI{\tempurl}


\bibitem[Wang et~al\mbox{.}(2022a)]%
        {Wang+al:22a}
\bibfield{author}{\bibinfo{person}{Yumeng Wang}, \bibinfo{person}{Lijun Lyu}, {and} \bibinfo{person}{Avishek Anand}.} \bibinfo{year}{2022}\natexlab{a}.
\newblock \showarticletitle{{BERT} Rankers are Brittle: {A} Study using Adversarial Document Perturbations}. In \bibinfo{booktitle}{\emph{Proceedings of {ICTIR}}}. \bibinfo{pages}{115--120}.
\newblock


\bibitem[Wei and Croft(2006)]%
        {Wei+Croft:06a}
\bibfield{author}{\bibinfo{person}{Xing Wei} {and} \bibinfo{person}{W.~Bruce Croft}.} \bibinfo{year}{2006}\natexlab{}.
\newblock \showarticletitle{{LDA}-Based document models for Ad-hoc retrieval}. In \bibinfo{booktitle}{\emph{Proceedings of SIGIR}}. \bibinfo{pages}{178--185}.
\newblock


\bibitem[White(2024)]%
        {White:25a}
\bibfield{author}{\bibinfo{person}{Ryen~W. White}.} \bibinfo{year}{2024}\natexlab{}.
\newblock \showarticletitle{Advancing the Search Frontier with {AI} Agents}.
\newblock \bibinfo{journal}{\emph{Commun. {ACM}}} \bibinfo{volume}{67}, \bibinfo{number}{9} (\bibinfo{year}{2024}), \bibinfo{pages}{54--65}.
\newblock


\bibitem[Wu et~al\mbox{.}(2022)]%
        {Wu+al:22a}
\bibfield{author}{\bibinfo{person}{Chen Wu}, \bibinfo{person}{Ruqing Zhang}, \bibinfo{person}{Jiafeng Guo}, \bibinfo{person}{Maarten de Rijke}, \bibinfo{person}{Yixing Fan}, {and} \bibinfo{person}{Xueqi Cheng}.} \bibinfo{year}{2022}\natexlab{}.
\newblock \bibinfo{title}{PRADA: Practical Black-Box Adversarial Attacks against Neural Ranking Models}.
\newblock
\newblock
\showeprint[arxiv]{2204.01321}


\bibitem[Wu et~al\mbox{.}(2010)]%
        {wu2010adapting}
\bibfield{author}{\bibinfo{person}{Qiang Wu}, \bibinfo{person}{Christopher~JC Burges}, \bibinfo{person}{Krysta~M Svore}, {and} \bibinfo{person}{Jianfeng Gao}.} \bibinfo{year}{2010}\natexlab{}.
\newblock \showarticletitle{Adapting boosting for information retrieval measures}.
\newblock \bibinfo{journal}{\emph{Information Retrieval}}  \bibinfo{volume}{13} (\bibinfo{year}{2010}), \bibinfo{pages}{254--270}.
\newblock


\bibitem[Xu and Croft(1996)]%
        {Xu+Croft:96a}
\bibfield{author}{\bibinfo{person}{Jinxi Xu} {and} \bibinfo{person}{W.~Bruce Croft}.} \bibinfo{year}{1996}\natexlab{}.
\newblock \showarticletitle{Query Expansion using Local and Global Document Analysis}. In \bibinfo{booktitle}{\emph{Proceedings of SIGIR}}. \bibinfo{pages}{4--11}.
\newblock


\bibitem[Zhai(2024)]%
        {Zhai:24a}
\bibfield{author}{\bibinfo{person}{ChengXiang Zhai}.} \bibinfo{year}{2024}\natexlab{}.
\newblock \showarticletitle{Large Language Models and Future of Information Retrieval: Opportunities and Challenges}. In \bibinfo{booktitle}{\emph{Proceedings of SIGIR}}. \bibinfo{pages}{481--490}.
\newblock


\bibitem[Zhang et~al\mbox{.}(2024)]%
        {Zhang+al:24a}
\bibfield{author}{\bibinfo{person}{Weinan Zhang}, \bibinfo{person}{Junwei Liao}, \bibinfo{person}{Ning Li}, {and} \bibinfo{person}{Kounianhua Du}.} \bibinfo{year}{2024}\natexlab{}.
\newblock \bibinfo{title}{Agentic Information Retrieval}.
\newblock
\newblock
\showeprint[arxiv]{2410.09713}~[cs.IR]


\bibitem[Zhu et~al\mbox{.}(2024)]%
        {Zhu+al:24a}
\bibfield{author}{\bibinfo{person}{Yutao Zhu}, \bibinfo{person}{Huaying Yuan}, \bibinfo{person}{Shuting Wang}, \bibinfo{person}{Jiongnan Liu}, \bibinfo{person}{Wenhan Liu}, \bibinfo{person}{Chenlong Deng}, \bibinfo{person}{Haonan Chen}, \bibinfo{person}{Zheng Liu}, \bibinfo{person}{Zhicheng Dou}, {and} \bibinfo{person}{Ji-Rong Wen}.} \bibinfo{year}{2024}\natexlab{}.
\newblock \bibinfo{title}{Large Language Models for Information Retrieval: A Survey}.
\newblock
\newblock
\showeprint[arxiv]{2308.07107}~[cs.CL]


\bibitem[Zhuang et~al\mbox{.}(2024)]%
        {zhuang2024setwise}
\bibfield{author}{\bibinfo{person}{Shengyao Zhuang}, \bibinfo{person}{Honglei Zhuang}, \bibinfo{person}{Bevan Koopman}, {and} \bibinfo{person}{Guido Zuccon}.} \bibinfo{year}{2024}\natexlab{}.
\newblock \showarticletitle{A Setwise Approach for Effective and Highly Efficient Zero-shot Ranking with Large Language Models}. In \bibinfo{booktitle}{\emph{Proceedings of the 47th International ACM SIGIR Conference on Research and Development in Information Retrieval (SIGIR '24)}}. \bibinfo{publisher}{Association for Computing Machinery}, \bibinfo{address}{New York, NY, USA}, \bibinfo{pages}{38--47}.
\newblock
\urldef\tempurl%
\url{https://doi.org/10.1145/3626772.3657813}
\showDOI{\tempurl}


\end{thebibliography}

\end{document}